\documentclass[journal, onecolumn, doublespacing, draftcls,12pt]{IEEEtran}
\usepackage[T1]{fontenc}
\usepackage{amssymb,amsfonts,amsthm,amsmath,bm}
\usepackage[color]{changebar}
\cbcolor{red}
\usepackage{amsmath}
\usepackage{geometry}
\usepackage[cmintegrals]{newtxmath}
\usepackage{url}
\usepackage{color}
\usepackage {graphicx}
\usepackage{multirow}
\usepackage{cite}
\usepackage{enumitem}
\usepackage{mathtools}
\usepackage{bigstrut} 
\usepackage{tikz}
\usepackage{pgfplots}
\usepackage[normalem]{ulem}
\theoremstyle{definition}
\newtheorem{thm}{Theorem}
\newtheorem{lem}{Lemma}
\def\x{{\mathbf x}}

\def\b{{\mathbf b}}
\def\u{{\mathbf u}}
\def\w{{\mathbf w}}

\def\e{{\mathbf e}}
\def\p{{\mathbf p}}
\def\y{{\mathbf y}}
\def\z{{\mathbf z}}

\def\x{{\mathbf x}}
\def\0{{\mathbf 0}}
\def\A{{\mathbf A}}
\def\I{{\mathbf I}}

\def\H{{\mathbf H}}
\def\P{{\mathbf P}}

\def\CN{{\mathcal{CN}}}

\def\N{{\mathcal{N}}}

\def\E{{\mathbb E}}

\begin{document}
\title{\Large Intelligent Reflecting Surface Assisted Beam Index-Modulation for Millimeter Wave Communication}
\author{Sarath Gopi and Sheetal Kalyani and Lajos Hanzo 	\thanks{\copyright 2020 IEEE.  Personal use of this material is permitted.  Permission from IEEE must be obtained for all other uses, in any current or future media, including reprinting/republishing this material for advertising or promotional purposes, creating new collective works, for resale or redistribution to servers or lists, or reuse of any copyrighted component of this work in other works.}} 
\maketitle
\vspace{-1cm}
\begin{abstract}
Millimeter wave communication is eminently suitable for high-rate
wireless systems, which may be beneficially amalgamated with
intelligent reflecting surfaces (IRS), while relying on beam-index
modulation. Explicitly, we propose three different architectures based
on IRSs for beam-index modulation in millimeter wave communication.
Our schemes are capable of eliminating the detrimental line-of-sight
blockage of millimeter wave frequencies.The
  schemes are termed as single-symbol beam index modulation,
  multi-symbol beam-index modulation and maximum-SNR single-symbol
  beam index modulation. The principle behind these is to embed the
  information both in classic QAM/PSK symbols and in the transmitter
  beam-pattern. Explicitly, we proposed to use a twin-IRS structure to
  construct a low-cost beam-index modulation scheme. We
conceive both the optimal maximum likelihood detector and a
low-complexity compressed sensing detector for the proposed
schemes. Finally, the schemes designed are evaluated through extensive
simulations and the results are compared to our analytical bounds.
\end{abstract}

    \section{Introduction}
   
 Next-generation systems are expected to satisfy substantially improved specifications. Furthermore, new solutions, such as the Internet of Things (IoT), massive machine type communications (MTC) also contribute to the escalating mobile data traffic, as predicted by the International Telecommunication Union (ITU) \cite{tariq2019speculative}. Hence researchers aim for increasing the degrees of design-freedom in support of these ambitious requirements. 
   \par 
  The $30-300~GHz$ so-called millimeter wave (mmWave) frequency band has substantial hitherto unexploited bandwidth resources for supporting Gigabit per seconds (Gb/s) data rates \cite{6732923, 6515173, rappaport2016spectrum}. For example, in indoor scenarios a data rate of upto $6.7~Gbps$ is achieved by the IEEE 802.11ad standard developed at $60$~GHz frequency~\cite{6392842}. This result has ignited research interest in this frequency range also for outdoor scenarios. In an early experiment, it has been shown that mmWave communication is capable of achieving a peak data rate of $1~Gbps$ in an outdoor environment for a communication range of upto $1.7~km$ at moderate Bit Error Rates (BERs)~\cite{6736750}.  This system used only $500$~MHz of bandwidth at $28$~GHz. Naturally, there are a number of propagation challenges to be overcome, since typically only line-of-sight (LOS) communication is possible at these frequencies, which also suffer from fading, significant absorption losses in the atmosphere and building-penetration losses~\cite{7999294}~\cite{7522613}. 
  \par 
 Furthermore,  researchers are also aiming for reducing both the power consumption and hardware cost. Intelligent Reflecting Surfaces (IRS) offer a viable solution for meeting these requirements~\cite{dai2019reconfigurable}. Explicitly, IRSs constitute passive reflecting surfaces equipped with integrated electronic circuits, which are capable of imposing carefully controlled amplitude and/or phase shifts on the incident signals \cite{qingqing2019towards, di2019smart, huang2019holographic}. The concept has been earlier proposed in \cite{6206517} and its employment as a phase-shifter has become popularized by \cite{7510962}. IRSs are eminently suitable for  energy-efficient solutions in a wide variety of applications, such as signal-to-noise-ratio (SNR) maximization \cite{yan2019passive}, rate-maximization \cite{yu2019miso}, \cite{han2019large}, for improving the energy efficiency  \cite{huang2019reconfigurable, huang2018energy}, for minimizing transmit power~\cite{8811733}, for providing secure communication \cite{chen2019intelligent}, multi-cell MIMO communication~\cite{pan2019multicell},~\cite{pan2019power}, over the air computation \cite{jiang2019over}, low latency mobile edge computing \cite{bai2019latency}, index modulation\cite{basar2020reconfigurable} and so on.
  \par 
  In \cite{8325484}, analog beamforming based beam-index modulation has been proposed as an extension of spatial modulation \cite{4382913, 6823072}. Inspired by these results, we conceive IRS assisted beam-index modulation for mmWave communications. Beamforming techniques have been exploited in mmWave communication for mitigating their path loss~\cite{7342886, 6955833, 6957140}, for achieving directional transmission~\cite{8468205, 7390101, 8962355}, for avoiding inter-carrier-interference~\cite{8978709} and also for safeguarding against eavesdroppers~\cite{8283647}. However, there is a paucity of contributions on beamforming-aided index modulation in IRS-assisted mmWave communication.  Our main contributions are:
 \begin{enumerate}
 	\item We propose IRS assisted beam index modulation for mmWave communication. Beamforming solutions proposed for mmWave frequencies tend to rely on either analog beamforming \cite{7870294, 8690621} or on hybrid techniques \cite{6736750, 7010533, 8901444, 7491314, 8924932, 8959381, 8643353, 8964330}.  In \cite{8883297} digital beamforming is proposed, which relies on complex hardware. As a remedy, IRS has been proposed for imposing phase shifts on the incident signal, which can be exploited for beamforming. As a further benefit, they are capable of  circumventing the predominantly LOS nature of mmWave propagation. Hence, our proposed scheme has at least three appealing features: it supports non-LOS communication at mmWave frequencies at a low cost, whilst conveying extra information via beam-index modulation. 
 	\item We propose three different architectures for IRS assisted beam-index modulation. The first is termed as single-symbol beam index modulation, where the information is carried both by classic QAM/PSK symbols and by the transmitter beam-pattern. This idea has also been extended for further improving the data rate in our Scheme 2, which is a multi-symbol beam-index modulation arrangement. In the third scheme, we provide an architecture for improving the SNR of the proposed beam-index modulation. 
 	\item The optimal maximum likelihood (ML) detector is derived for the schemes conceived. Additionally, a low complexity compressed sensing assisted detector is also developed. 
 	\item An upper bound of the average BER is obtained for the optimal ML detector. Finally, the proposed scheme is evaluated through extensive simulations and its performance is compared to the theoretically obtained bound.
 \end{enumerate}
\begin{table}[b]
	\begin{center}
	\begin{tabular}{|l|l|l|l|l|l|l|l|}
	\hline
	& Our Scheme & \cite{8325484}-2018 & \cite{yan2019passive}-2019 & \cite{canbilen2020reconfigurable}-2020 & \cite{basar2020reconfigurable} -2020   &\cite{8964330} -2020 & \cite{yang2020mimo}-2020\\
	\hline
	Multi IRS assisted multihop & $\checkmark$ & & & & & & $\checkmark$ \\
	\hline
	mmWave &$\checkmark$ & & & & & $\checkmark$ & $\checkmark$ \\
	\hline
	non-LOS &$\checkmark$ & $\checkmark$& & $\checkmark$& $\checkmark$  & $\checkmark$& \\
	\hline 
	Beam Index Modulation &$\checkmark$& $\checkmark$ & & & & & \\
	\hline
	SNR Optimization &$\checkmark$ & &$\checkmark$& $\checkmark$& $\checkmark$&   &\\
	\hline
	Beamformer Gain &$\checkmark$ & & & &  & $\checkmark$ & $\checkmark$ \\
	\hline 
	Rician Channel Model &$\checkmark$ & & & & &$\checkmark$ & \\
	\hline BER analysis based on&$\checkmark$ & & & & & &\\
	Non-Gaussian Approximation & & & & & & & \\
	\hline
	\end{tabular}
		\caption{Comparison of the proposed scheme with similar ideas.}
			\end{center}
	\label{CompTable}
\end{table}
In Table \ref{CompTable}, we provide a bold
  summary and contrast our new contributions to the seminal
  literature. The key contribution of our scheme is a unique twin-IRS
  architecture, in which one of the IRSs can be positioned farther
  away from the transmitter. IM on this IRS is activated wirelessly
  using the other IRS. This architecture benefits in terms of
  accomplishing non-LOS communication by two LOS paths, additionally
  achieving a substantial beamformer gain and hence an SNR gain. 
The rest of the paper is organized as follows. Section \ref{p4prop}
details the proposed IRS assisted beam-index modulation schemes. The
implementation aspects and parameter design of the schemes are
detailed in Section \ref{p4sysimp}, while our detectors are developed
in Section \ref{p4det}. In Section \ref{p4ana}, the error analysis of
the proposed scheme is provided. Our simulation results are given in
Section \ref{p4sym} and we conclude in Section \ref{p4concl}.  \\ {\bf
  {Notations:}} Throughout the paper, unless otherwise specified, bold
lower case and bold upper case letters are used to represent vectors
and matrices, respectively.  $\A^H$, $\text{Tr}\{\A\}$ and
$\lambda_{min}(\A)$ represents the hermitian, trace and minimum eigen
value of $\A$, respectively. $\|.\|$ stands for $L_2$- norm. $|a|$ and
$\mathcal{R}e\{a\}$ is the absolute and real value of scalar $a$,
respectively. $\I$ is the identity matrix of appropriate
dimension. $\CN(\mu, \mathbf{C})$ represents the complex Gaussian
distribution with mean vector $\mu$ and covariance matrix
$\mathbf{C}$. $\lfloor b \rfloor$ is the largest integer not greater
than $b$. $\Gamma(.)$ is the $\Gamma$ -function, i.e., $\Gamma(z) =
\int_{0}^{\infty}x^{z-1}e^{-x}dx$ and for integer $z$, $\Gamma(z) =
(z-1)!$ and $\Gamma(a,b)$ is the Gamma distribution with $a$ and $b$
are shape and rate parameters, respectively.
   \section{Proposed IRS Assisted Beam-Index Modulation Schemes}
   \label{p4prop}
\begin{figure}[b]
   	\centering
   	\includegraphics[width = 0.8\textwidth]{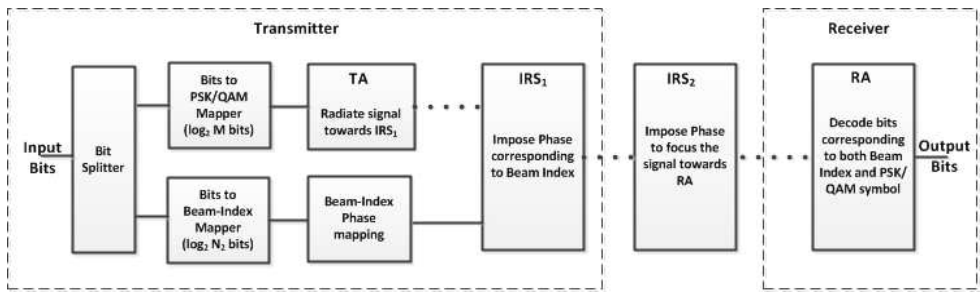}
   	\caption{Proposed IRS assisted beam-index modulation scheme.  Solid and dotted lines indicate wired and wireless links, respectively. }
   	\label{p4fig0}
   \end{figure}

\begin{figure}
   	\centering
   	\includegraphics[width = 0.50\textwidth]{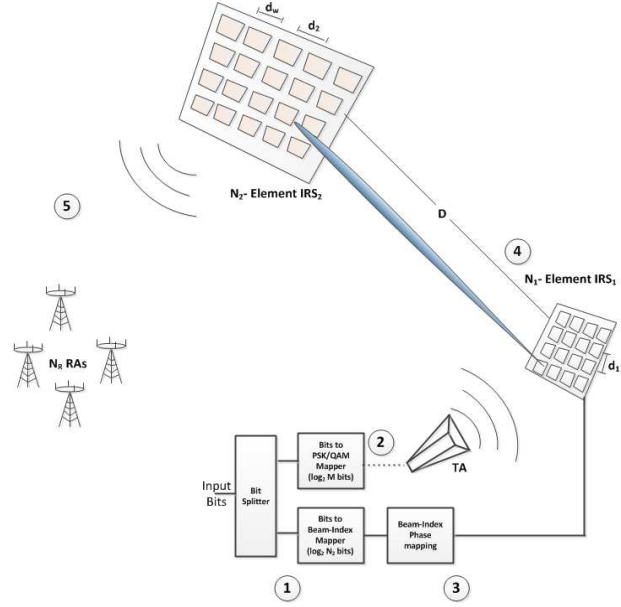}
   	\caption{Architecture of single-symbol beam-index modulation.}
   	\label{p4fig1}
   \end{figure}

   We propose three different IRS assisted beam-index modulation
   schemes. All these schemes are single input multiple output (SIMO)
   arrangements, containing a transmitter antenna (TA), two sets of
   IRSs and $N_R$ receiver antennas (RAs). Each IRS has one or more
   reflecting surfaces (RS) and each RS has many elements. The first
   IRS, namely $IRS_1$, can be directly accessed by the transmitter
   and it is used for selectively activating the elements in the
   second IRS, i.e. in $IRS_2$. The block
     diagram of the proposed scheme is sketched in Fig.~\ref{p4fig0},
     which is further elaborated on using Fig.~\ref{p4fig1}. The steps
     from $\textcircled{1}$ to $\textcircled{5}$ in the
     Fig.~\ref{p4fig1} is detailed below.
   \begin{enumerate}
   	\item The incoming bit sequence is split into two groups. The first group is used for selecting the classic PSK/QAM symbols, while the second set is used for beam-index modulation.
   	\item The TA and $IRS_1$ are kept close to each other. They have both wired and wireless connections. Based on the first group of bits, an appropriate PSK/QAM symbol ($s$) is selected at the transmitter, which is transmitted wirelessly to each RS in $IRS_1$. 		
   	\item The wired connection is used for mapping the second group of bits onto beam-index modulation. These bits are converted to the appropriate phase vector, which are then forwarded to the elements of the RSs in $IRS_1$.
   	\item Based on the received phase vector, the elements in $IRS_1$ impose the required phase shift on the incident signal, which are then forwarded to $IRS_2$. The phase is specifically adjusted for ensuring that only the desired elements in $IRS_2$ receive the signal. This specific selection is determined based on the information bits reserved for beam-index modulation.
   	\item Each element of $IRS_2$ induce a constant phase  to reflect the signal towards the RAs. This is captured by the RAs. The information detected at the RAs includes both the conventional PSK/QAM symbols and the specific element indices of $IRS_2$, which reflect the symbols. This is done jointly by $N_R$ RAs.
   \end{enumerate}
For detailing the schemes, we will make the following assumptions. 
   \begin{enumerate}
   	\item The channel between the TA and
          $IRS_1$ may be deemed to be a low-noise AWGN channel. A
          conventional horn antenna having a few centimetre length can
          be used as the TA, where $IRS_1$ is positioned, say $4$-$5$
          m away from the TA~\cite{sethi2013millimeter}.  Hence, the
          distance satisfies the far field condition, albeit this
          is not actually necessary, since the position of the TA is
          perfectly known at $IRS_1$, hence near-perfect delay
          compensation can be arranged for this location. The
          TA is designed in such a way that the signal is pointed
          exactly towards $IRS_1$. Finally, $IRS_1$, which is a
          passive device, introduces a phase shift and reflects the
          signal towards $IRS_2$. Hence, the only source of noise,
          that can affect the signal is the one, which is added at
          $IRS_1$, and this is negligible.
   	\item There is only LOS communication between $IRS_1$ and
          $IRS_2$. Typically, a Rician
          channel model is used for modelling IRS-assisted
          communication systems~\cite{zhou2020framework}. However, for
          the proposed scheme the elements in $IRS_1$ adjust the phase
          in such a way that it forms a directional beam and only the
          specifically selected elements of $IRS_2$ receive the
          signal.  Moreover, $IRS_2$, which is close to $IRS_1$, is
          carefully positioned for ensuring that there is no blockage
          between $IRS_1$ and $IRS_2$.  Hence, the channel between
          $IRS_1$ and $IRS_2$ is assumed to be an AWGN channel.
   	\item Between $IRS_2$ and the receiver, we have a Rician channel model.
   \end{enumerate}
 A beneficial application of the proposed
   architectures can be found in the Internet of Things(IoT), where
   the desired information has to be collected by sensors and
   delivered to either a distant server or to a user. The various applications
   include smart homes, industrial and environmental monitoring,
   building and home automation etc. Let us assume that the
   information collected from a home or an industrial cite should be
   communicated to a BS, from where the information can be
   communicated to the destination through the cellular network. In
   this case, the TA and $IRS_1$, which only belong to the specific
   user, can be placed in the terrace of the home or at the industrial
   cite. The $IRS_2$ can be situated at the top of a tall building in
   the vicinity, which can be shared among many such users, who have
   orthogonal resources. The details of the schemes are given
 below.

   \subsection{Scheme 1: Single-Symbol Beam-Index Modulation}
   This scheme is shown in Fig.~\ref{p4fig1}. In this scheme, both the IRSs  have only a single RS. $IRS_1$ is directly connected to the transmitter, whereas $IRS_2$ is kept at a distance, say $D$, from the first IRS. Let $IRS_1$ be a $\left(N_{1H} \times N_{1W}\right)$ element array, while $IRS_2$ be an $\left(N_{2H} \times N_{2W}\right)$ array and let $N_1 = N_{1H}N_{1W}$ and $N_2 = N_{2H}N_{2W}$ be the total number of elements in $IRS_1$ and $IRS_2$, respectively. The TA sends the symbols to $IRS_1$, where each element applies a specific phase shift to the incident wave so that only one of the elements in $IRS_2$ receives the signal. Hence, in this scheme the total number of bits per channel use (bpcu) is $\log_2 M + \lfloor \log_2 N_2 \rfloor$, where the first term corresponds to the QAM/PSK symbols, while the second term corresponds to the selection of the element in $IRS_2$.   Finally, $IRS_2$ reflects the signal and it is received at the RAs. 
   \par
   Let $s$ be the transmitted symbol.  The symbol received at $IRS_1$ is $s+w_1$, where $w_1 \sim \CN(0,\sigma_1^2)$. However, under Assumption 1, we have $\sigma_1^2  \approx 0$ and the contribution $w_1$ can be discarded. Therefore, the vector received at $IRS_2$ is: 
   \begin{align}
   \x_2 = \b s + \w_2,
   \label{p4eqn1}
   \end{align}
     
  where $\w_2 \sim \CN(0,\sigma_2^2)$ under Assumption 2) and $\b$ is an $N_2 \times 1$ vector.  Ideally, $\b$ should have only a single non-zero entry corresponding to the index of the beam (or equivalently corresponding to the selected element in $IRS_2$). However, this will not happen in practice, since a finite power will be dispersed on other directions also and this power distribution depends on the beampattern. The vector at the receiver can be written as:
   \begin{align}
   \y &= \H\Theta\x_2 + \w_R \nonumber \\
   &= \H\Theta\b s + \w,
   \label{p4eqn2}
   \end{align}
   where $\H$ is the $N_R \times N_2$ channel matrix as defined under
   Assumption 3, $\Theta$ is an $\left(N_2 \times N_2 \right)$
   diagonal matrix of phase shifts given by the elements in $IRS_2$
   and $\w_R \sim \CN(0,\sigma_R^2)$. Note that $\w= \H \Theta\w_2 +
   \w_R$ is the additive noise component having a distribution of
   $\CN\left(0, \Sigma\right)$, where $\Sigma = \H\H^H\sigma_2^2 +
   \sigma_R^2\I$. Finally, the receiver has to detect both $\b$ and
   $s$ from $\y$ to decode the transmitted bits. The detection schemes
   will be discussed in Section \ref{p4det}.  \par This scheme predominantly uses only a fraction of elements of the second reflecting surface, instead of exploiting
     all of them to improve the attainable beamformer gain. However,
     it should be noted that most of the transmitted energy is
     focussed on the intended elements, while the power impinging on
     all other elements is negligibly small, since beamforming is
     used in the first stage. Hence, with the aid of the proposed
     scheme, we will get the dual advantages of both a beneficial
     beamforming gain and the additional advantage of an increased
     data rate.
   \subsection{Scheme 2: Multi-Symbol Beam-index Modulation}
\begin{figure}
   \centering
   \includegraphics[width = 0.50\textwidth]{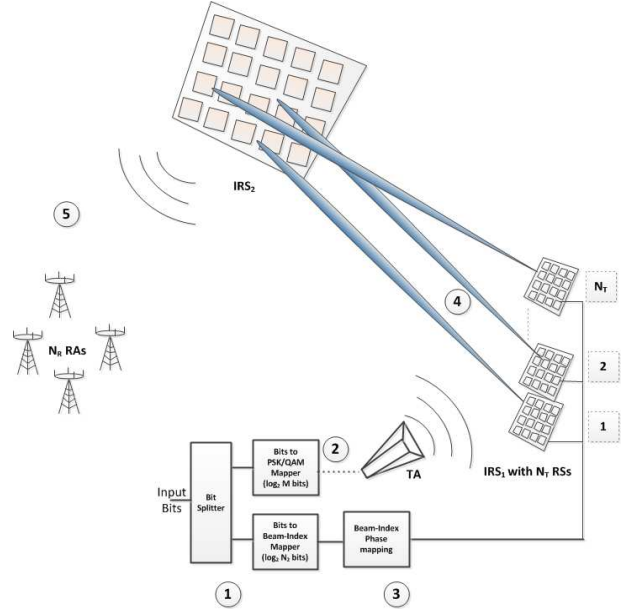}
   \caption{Architecture of multi-symbol beam-index modulation. Contrast to Fig. ~\ref{p4fig1}, there are $N_T$ RSs in $IRS_1$ in this case.}
   \label{p4fig2}
\end{figure}

   In the second scheme, the first scheme is extended to multi-symbol communication. The architecture is shown in Fig.~\ref{p4fig2}. In this case, there are $N_T$ RSs in $IRS_1$ contrast to a single RS in Scheme 1. The modulator identifies $N_T$ different phase-vectors depending on the bit sequence corresponding to the beam-index modulation and each vector is fed to different RSs in $IRS_1$. Therefore, $N_T$ RSs focus the conventional QAM/PSK symbol onto $N_T$ different elements of $IRS_2$. Hence, in this case, the total number of bpcu is $\log_2 M + \left\lfloor \log_2 \binom{N_2}{N_T} \right\rfloor$. Therefore, this scheme provides a higher data rate than scheme 1. The choice of the elements to be activated can be organized using a look up table method or the combinatoric approach \cite{6587554}, \cite{8737925}.
\par 
Mathematically, this scheme can be represented using Equations (\ref{p4eqn1}) and (\ref{p4eqn2}). However, the difference is that in this case, ideally there will be $N_T$ non-zero entries in $\b$.

   \subsection{Scheme 3: Maximum-SNR Single-Symbol Beam-Index Modulation}
\begin{figure}
    	\centering
    	\includegraphics[width = 0.50\textwidth]{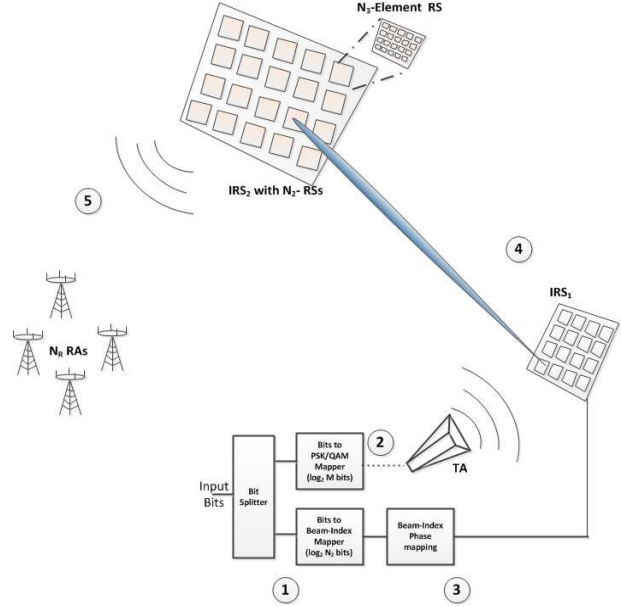}
    	\caption{Architecture of the Maximum-SNR Single-Symbol Beam-Index Modulation. Contrast to Fig.~\ref{p4fig1}, there are $N_3$ elements in each RS of $IRS_2$, which impose phase shift on the incident signal to maximize SNR at the RAs.}
    	\label{p4fig3}
    \end{figure}

    Fig.~\ref{p4fig3} shows the architecture of this scheme. This is similar to Scheme 1, except that in this case each element of $IRS_2$ is replaced by an RS having $N_3$ elements. Hence, there will be a total of $N_2N_3$ elements in $IRS_2$. Both the TA and $IRS_1$ function in the same way as in the case of single-symbol beam-index modulation. Hence, the signal received at $IRS_2$ can be written using (\ref{p4eqn1}).  However, in contrast to the other two cases, here the elements in $IRS_2$ apply a phase shift to the incident signal.  The phase shift in $IRS_2$ is adjusted in such a way that the SNR at the receiver is maximized.  
    \par 
  
  Note that in (\ref{p4eqn2}) $\H$ and $\Theta$ is an $\left(N_R \times N_2N_3\right)$ and $\left(N_2N_3 \times N_2N_3\right)$ matrix, respectively in this case. Furthermore, $\Theta$ is not a constant matrix, but depends on $\H$. The overall SNR in this case is defined as:
    \begin{align}
    \text{SNR} = \frac{\|\H\Theta\b s\|^2}{\text{Var}(\|\w\|)},
    \label{p4eqn3a}
    \end{align} 
    where the denominator is the variance of the norm of the vector $\w$. The SNR can be maximized by maximizing the numerator of (\ref{p4eqn3a}), since the denominator is independent of $\Theta$, which is the maximization variable.  Let $\theta_{1:N_2N_3}$ represents the entries of the diagonal of $\Theta$. Hence, the SNR maximization can be written as: 
   \vspace{-0.3cm}
    \begin{align}
    \underset{\theta_l}{\max}~\|\H\Theta \b s\|^2~s.t.~|\theta_l| =1~, \forall~l = 1, 2, ..., N_2N_3.
    \label{p4eqn4}
    \end{align}
    \par
  However, the above optimization problem has the following challenges.  $IRS_2$ is a passive device and it may not be practical to solve a complex optimization problem there. Hence, the optimization should ideally be carried out at transmitter or receiver and the resultant information has to be communicated to $IRS_2$. Therefore, if the optimization depends on the data to be transmitted ($\b s$), $\Theta$ has to be updated in every time slot, which is a substantial communication overhead. Hence, the optimization should preferably only depend on either an average value of $\b s$ or indeed ideally should be independent of  it. Accordingly, we will propose the following solutions for (\ref{p4eqn4}).
    
    \subsubsection{Solution 1}
    This solution is based on the assumption that an ideal beam pattern exists, i.e., all elements in the selected RS of $IRS_2$ receives the same power, while all other elements receive no power. Without loss of generality, let this constant be $1$. Hence, the optimization function in (\ref{p4eqn4}) can be written as:
    \begin{align}
    \|\H\Theta \b s\|^2 = \|\H\Theta \textbf{1}\|^2 = \|\H\bm{\theta} \|^2,
    \label{p4eqn4a}
    \end{align}
 where $\textbf{1}$ is a vector of 1s and $\bm{\theta}$ is a vector formed from the diagonal elements $\theta_{1:N_2N_3}$ of $\Theta$. Hence, the maximization problem (\ref{p4eqn4}) becomes:
    \begin{align}
    \underset{{\theta}_l}{\max}~{\bm{\theta}}^H\H^H\H{\bm{\theta}}~s.t.~|{{\theta}}_l| =1~, \forall~l = (\hat{I}-1)N_3+1, ..., \hat{I}N_3,
    \label{p4eqn4b}
    \end{align}
where $\hat{I}$ is the specifically selected RS in $IRS_2$. Let $\theta_l = e^{j\alpha_l}$, since $|\theta_l| =1$. Bearing this in mind and noting that $\H^H\H$ is a Hermitian matrix, (\ref{p4eqn4}) is reformulated as the following unconstrained optimization problem.
     \begin{align}
     \underset{\alpha_l}{\max}~ \sum_{i=(\hat{I}-1)N_3+1}^{\hat{I}N_3}\sum_{j=(\hat{I}-1)N_3+1}^{\hat{I}N_3}  \mathcal{R}e\left\{e^{j\left(\alpha_i-\alpha_j\right)}\left(\H^H\H\right)_{ij}\right\},
     \label{p4eqn4c}
     \end{align}
  where $\left(\H^H\H\right)_{ij}$ is the $(i,j)^{th}$ element of $\H^H\H$. Since (\ref{p4eqn4c}) is not a concave function, it can only be solved using some iterative technique for finding its local maximum. 
     \subsubsection{Solution 2}
     \label{p4OptSol2}
    This solution relies on the assumption that $N_R \ge N_3$, i.e. there are more number of RAs than the number of elements in the RS of $IRS_2$. In order to develop the solution, let us  state and prove Lemma \ref{p4_lem_2}.
      \begin{lem}
      	\label{p4_lem_2}
     Let $\H_Q$ and $\Theta_Q$ be the $\left(N_R \times N_3\right)$ and $\left(N_3 \times N_3\right)$ sub-matrices of $\H$ and $\Theta$ corresponding to the selected RS, respectively and let $\b_Q$ be the corresponding sub-vector of $\b$. If $N_R \ge N_3$,  with probability 1, the bound	
      	\begin{align}
      	 \|\H_Q\Theta_Q \b_Q s\|^2 \ge  \lambda_{min}\left(\Theta_Q^H\H_Q^H\H_Q\Theta_Q \right) \text{Tr} \left\{(\b_Q s) (\b_Q s)^H\right\}
      	\label{p4eqnlem2_1}
      	\end{align}    		
      	is non-trivial, which equivalently leads to $\lambda_{min}\left(\Theta_Q^H\H_Q^H\H_Q\Theta \right) > 0$.
          	\begin{proof}
      	 	See Appendix \ref{p4_lem_2_proof} for proof.
      	 \end{proof}
      \end{lem}
 Lemma \ref{p4_lem_2} can be used  for solving the optimization problem (\ref{p4eqn4}). The idea is to maximize the non-trivial lower bound instead of the actual function. Hence, the optimization problem (\ref{p4eqn4}) becomes:
   \begin{align}
   \underset{\theta_l}{\max}~\lambda_{min}\left(\Theta_Q^H\H_Q^H\H_Q\Theta \right)
  ~~s.t.~|\theta_l| =1~, \forall~l = (\hat{I}-1)N_3+1, ..., \hat{I}N_3.
     \label{p4eqn4e}
   \end{align}
We know that $\lambda_{min}(\A) = \underset{\|\z\|=1}{\min} ~\|\A\z\|$ \cite[Eq. 7.5.4]{meyer2000matrix}. Therefore (\ref{p4eqn4e}) can be rewritten as: 
 	 \begin{align}
 	 &\underset{\theta_l}{\max}~ \underset{\z}{\min}~ \|\Theta_Q^H\H_Q^H\H_Q\Theta\z\|~\nonumber \\
 	  &~~~~s.t.~\|\z\| = 1,~~~|\theta_l| =1~, \forall~l = (\hat{I}-1)N_3+1, ..., \hat{I}N_3,
 	 \label{p4eqn4f}
 	 \end{align}  
	where (\ref{p4eqn4f}) is a constrained non-linear minimax optimization problem. This can be solved directly \cite{ deb2012optimization, osborne1969algorithm}. Alternatively, it can be converted into a non-linear maximization problem by introducing an additional variable and then solved using standard techniques. 
     \par 
It should be noted that in both solutions of the SNR maximization problem, $\hat{I}$, i.e. the selected data dependent RS of $IRS_2$ that has to be optimized, depends on the information bits. In order to avoid the dependence of optimization on the information bits, each RS is optimized separately whenever there is considerable change in the channel.  The optimized phase information is passed to $IRS_2$, which applies phase shifts to all elements instead of the selected RS. This scheme can be extended to the case of multi-symbol beam-index modulation (Scheme 2), where there will be $N_T$ RSs in $IRS_1$, which activate $N_T$ RSs in $IRS_2$. Finally, all the activated RSs in $IRS_2$ can apply a phase shifts for improving the SNR. Thus the scheme will have both an improved data rate and improved SNR. Practically, the optimal phase shifts have to be estimated at the receiver and then communicated to $IRS_2$ whenever there is significant change in the channel characteristics. 
   
\section{Implementation of Beam-Index Modulation} 
\label{p4sysimp}
The principle behind the proposed beam-index modulation is the data-dependent activation of the elements in $IRS_2$. This is achieved by appropriately choosing the phase shifts applied by the elements in $IRS_1$. In order to estimate the phase shifts, it is assumed that there is only LOS communication between $IRS_1$ and $IRS_2$. Therefore, the phase shifts only depend on the geometry of the pair of IRSs. The estimation of phase shifts is detailed below. 
\par 
Let the centre of $IRS_1$ be the origin co-ordinate $(0,0,0)$ and $\P$ be the position vector of elements of $IRS_1$.  Let the $n^{th}$ element of $IRS_2$ be activated by $IRS_1$ according to the input bit sequence and let $(\phi_n^h, \phi_n^v)$ represents the azimuth and the elevation angle pair for this element with respect to the origin. Then, the phase-vector to be given by the elements of $IRS_1$ to choose the $n^{th}$ beam is  $\psi = 2\pi f\tau$, where $\tau = \frac{\P^H\u_n}{c}$ with $c$ and $f$ being the speed of the light and the carrier frequency, and $\u_n = \left(\sin \phi_n^h\cos \phi_n^v  ~\cos\phi_n^h\cos \phi_n^v ~ \sin \phi_n^v \right)$. Finally, in the case of multi-bit beam-index modulation, these phase shifts have to be calculated for each of IRSs according to the input bit sequences. 
\par 
Additionally, if the IRS elements can modify the amplitude of the incident signal along with the phase, one can modify the relative weighting of each element. Since IRSs constitute passive devices, amplification may be difficult to achieve and will not be a cost effective solution. However, attenuation can be readily applied to the incident signal \cite{7744497}. The attenuation can be adjusted in such a way that it acts as a window function for the beamforming and the beam pattern can be accordingly modified. This can be used to reduce signal received by unwanted elements in $IRS_2$. In Section \ref{p4sysdesEx}, the design of two IRSs is detailed. 

\subsection{Parameter Design}
\label{p4sysdesEx}
The parameters to be designed are the number of elements and the corresponding inter-element spacing in $IRS_1$, as well as in $IRS_2$ and the distance between two IRSs.  Let $\lambda$ be the wavelength corresponding to the highest frequency of operation. We will fix the design parameters as follows~\cite{van2004optimum}.
\begin{enumerate}
\item \emph{Inter-element spacing in $IRS_1$ ($d_1$):} The elements in each RSs of $IRS_1$ should be spaced at distances $\le  \frac{\lambda}{2}$. For example, if the maximum operating frequency is $f_{max} = 60 ~GHz$, then the spacing between the elements in $IRS_1$ is $d_1 \le 2.5~mm$. Now, if $IRS_1$ is a $100 \times 100$ element system, then its dimension is going to be as compact as $0.25 \times 0.25~m$.  This spacing is important for avoiding grating lobes in the beams formed using $IRS_1$~\cite[eq. 2.117]{van2004optimum}. 
	\item \emph{Distance between two IRSs ($D$):} The distance ($D$) between $IRS_1$ and $IRS_2$ should meet the far field condition of $D > \frac{2L^2}{\lambda}$, where $L$ is the length of the RS~\cite[pp. 32]{balanis2016antenna}. In the above example $D > 25 ~m$. If this condition is met, it can be assumed that the wave front travelling from $IRS_1$ to $IRS_2$ is planar, so that the phase shifts can be computed as detailed in Section \ref{p4sysimp}.
	\item \emph{Inter element spacing in $IRS_2$ ($d_2$):} The width of each element in $IRS_2$ ($d_w$) should be less than $D\theta_{BW}$, where $\theta_{BW}$ is the beam-width of $IRS_1$ and the separation $d_2$ between elements in $IRS_2$ should be higher than this value. These conditions ensure that the intended element and \emph{only the intended element} receives the signal reflected by $IRS_1$. For a rectangular window, the approximate beam-width is $\theta_{BW} \approx \frac{50\lambda}{L}$~\cite[Eq. (2.100)]{van2004optimum}. In the example we have considered $\theta_{BW} \approx1^0$ corresponding to the minimum frequency. Therefore $d_w < 48~cm$ and $d_2 > 48~cm$ for $D = 25~ m$. Now, if  $N_{1H}= N_{1W}= 8$, $IRS_2$ has an approximate dimension of $4~ m \times 4~ m$. 
	\item \emph{Number of elements in $IRS_1$ ($N_1$):} The number
          of elements in $IRS_1$ determines the length $L$ of the
          array and its beam-width $\theta_{BW}$, where these
          parameter decide the spacing between two IRSs and the
          inter-element spacing in $IRS_2$. A larger value of $N_1$ with accurate phase
            shifting and a larger value of $d_2$ direct the beam to
            the desired direction, hence reducing the potential beam
            misalignment problem of our beam index modulation scheme.
	\item \emph{Number of elements in $IRS_2$ ($N_2$):} This determines the data rate of the system. A large value of $N_2$ gives a higher data rate. However, this will make the size of $IRS_2$ large. Hence, $N_2$ is restricted by the maximum affordable array dimension. 
\end{enumerate}

 \section{Detector}
 \label{p4det}
 The detector has to recover the bits embedded both into the QAM/PSK symbol and the TA activation pattern in $IRS_2$. Explicitly, it has to detect $s$ and $\b$ from $\y$ in (\ref{p4eqn2}). Let $\x = \b s$ and $\A = \H\Theta$. Now, (\ref{p4eqn2}) can be written as:
 \begin{align}
 \y = \A\x + \w.
 \label{p4eqn5}
 \end{align}
 It is assumed that $\A$ is known at the receiver. 
 \subsection{ Optimal Detector}
 \label{p4optimdet}
 We first derive the optimal ML detector for the single symbol cases, i.e. for Scheme 1 and Scheme 3. Then we extent it to the multi symbol case of Scheme 2. 
\subsubsection{Single-Symbol Schemes} 
Consider the vector $\x$ in (\ref{p4eqn5}). Ideally in single-symbol schemes only one of the entries in $\x$ should be a non-zero value, since only one element receives the symbol. However, this will not be the case in practice, since the beamformer will introduce a non-zero power also in directions other than the required one. Hence, practically more than one element of $IRS_2$ receives the symbol. However, the power in the undesired beam-indices is much lower than that in the intended index and these powers depend on the beampattern of $IRS_1$.  The data-dependent beam-index changes can be represented approximately by a beampattern rotation. Hence, (\ref{p4eqn5}) can be written as:
%

  \begin{align}
  \y = \A \Pi_{\p} + \w,
  \label{p4eqn5a}
  \end{align}
  where $\p$ represents the vector of powers in the various indices of $\x$ and $\Pi_{\p}$ represents a particular permutation of the power pattern. Therefore, in order to identify the beam-index, we have to identify the power pattern permutation $\Pi_{\p}$. Now, $\y \sim \CN(\A \Pi_{\p},\Sigma)$. Hence, the ML detector of this problem is formulated as:
  \begin{align}
  \underset{\Pi_{\p}}{\min} &~ \left(\y - \A \Pi_{\p}\right)^H\Sigma^{-1} \left(\y - \A \Pi_{\p}\right) \nonumber \\
  \implies \underset{\Pi_{\p}}{\max} &~ \mathcal{R}e\left\{\left(\y -\frac{1}{2} \A \Pi_{\p}\right)^H\Sigma^{-1}\A \Pi_{\p} \right\}.
  \label{p4eqn6}
  \end{align}
   In general, the search problem (\ref{p4eqn6}) is NP-hard. However, in our case, there are only $N_2M$ different patterns corresponding to $N_2$ different beam indices and $M$ QAM/PSK symbols. Hence, a moderate-complexity search will give the optimal solution to the ML problem (\ref{p4eqn6}).  
   \subsubsection{Multi-Symbol Scheme}
   The ML detector (\ref{p4eqn6}) is also suitable for  multi-symbol case. However, in this case, since there are $N_T$ desired beam-indices at a time, which interact with each other and thereby produce a large number of possible combinations $\Pi_{\p}$. Explicitly, $\binom{N_2}{N_T}M$ different patterns hypothesis must be tested for $N_T$ RSs in $IRS_1$. Hence the ML detector may no longer be a computationally  attractable solution. Hence, in Section \ref{p4csdet}, we will be proposing a suboptimal compressed sensing (CS) aided detector, which can be used for any of the proposed schemes at a lower computational complexity.
 \subsection{Suboptimal Compressed Sensing Detector}
 \label{p4csdet}
 The transmitted vector $\x$ in (\ref{p4eqn5}) is sparse, when the number of active elements (i.e., elements that receive the symbol) is much less than the total number of elements in $IRS_2$. Therefore, one can use an efficient sparse reconstruction algorithm \cite{342465, needell2009cosamp} for identifying the non-zero components in $\x$, which can be used to estimate $\b$. However, it should be noted that for the successful recovery of the sparse vector $\x$, there should be a sufficient number of measurements. This can be either achieved by having a sufficient number of RAs ($N_R$ should be sufficiently large) or taking multiple measurements, which would naturally reduce the data rate. Finally, $s$ can be obtained from the estimated $\b$ as:
 \begin{align}
 \hat{s} = \underset{s \in \mathcal{M}}{\min}~ \| \y - \b s\|^2,
 \label{p4eqn6a}
 \end{align}
 where $\mathcal{M}$ is the constellation used.
\subsection{Complexity}
The optimal ML detector has to compute
  (\ref{p4eqn6}) for all possible combinations, which requires
  approximately on the order of $(N_R^3 + N_RN_2)$
  multiplications. This has to be done for each possible symbol. For
  the multi-symbol case, there are $\binom{N_2}{N_T}M$ possible
  symbols. Hence, the total computational complexity is approximately
  on the order of $\binom{N_2}{N_T}M(N_R^3 + N_RN_2)$, which reduces
  to the order of $N_2M(N_R^3 + N_RN_2)$ for single-symbol
  cases. On the other hand, if a greedy compressed sensing based
  suboptimal algorithm is used, the complexity will be reduced to the
  order of $N_2N_RN_T$, which is much lower than that of the optimal
  ML detector.
\section{Average Bit Error Rate Analysis}
\label{p4ana}
In this section, we will estimate an upper bound for the average bit error rate (BER) of the optimal ML detector of Section \ref{p4optimdet}. Let $\Pr\left\{ \Pi_{\p}^i \rightarrow \Pi_{\p}^j \right\}$ represent the probability that the pattern $\Pi_{\p}^i$ is identified as $\Pi_{\p}^j$ and $\nu_{i,j}$ represent the number of bits in error between the two permutations $\Pi_{\p}^i$ and $\Pi_{\p}^j$. Then the average BER is formulated as:
\begin{align}
\hat{\text{BER}} = \sum_{i=1}^{\Omega}\sum_{\underset{j \ne i}{j =1}}^{\Omega}\frac{\nu_{i,j}}{n_b\Omega}\Pr\left\{ \Pi_{\p}^i \rightarrow \Pi_{\p}^j \right\}, 
\label{p4eqn7a_1}
\end{align}
where $n_b$ is the total number of bits per channel use and $\Omega$ is the total number of possible permutations. Equation (\ref{p4eqn7a_1}) assumes that all permutations are equally likely. The probability of symbol error $\Pr\left\{ \Pi_{\p}^i \rightarrow \Pi_{\p}^j \right\} $ in (\ref{p4eqn7a_1}) can be found as follows. When $\Pi_{\p}^i$ is transmitted, the detector identifies $\Pi_{\p}^j$ as the transmitted symbol  based on:
\begin{align}
\underset{k}{\arg\max} ~\mathcal{R}e\left\{\left(\y - \frac{1}{2} \A \Pi_{\p}^k\right)^H\Sigma^{-1}\A \Pi_{\p}^k \right\} = j,
\label{p4eqn7b_1}
\end{align}
where $\y$ is given in (\ref{p4eqn5a}) in conjunction with $\Pi_{\p}= \Pi_{\p}^i$. Let us define $r_k = \mathcal{R}e\left\{\left(\y - \frac{1}{2} \A \Pi_{\p}^k\right)^H\Sigma^{-1}\A \Pi_{\p}^k \right\}$. Hence, we have
\begin{align}
\Pr\left\{ \Pi_{\p}^i \rightarrow \Pi_{\p}^j \right\} = \Pr\left\{\underset{k \ne j}{\bigcap} \left(r_j > r_k\right) \right\}.
\label{p4eqn7c_1}
\end{align}
The computation of the probability of intersection of the event in (\ref{p4eqn7c_1}) is very difficult. Hence, it is bounded using Fretchet's inequality \cite{frechet1935generalisation} as follows:
\begin{align}
 \Pr\left\{\underset{k \ne j}{\bigcap} \left(r_j > r_k\right) \right\}  \le \underset{k}{\min}~\Pr\left\{ r_j > r_k \right\}.
\label{p4eqn7d_1}
\end{align}
In order to estimate the bound, the probabilities of $\Pr\left\{ r_j > r_k \right\}$ have to be calculated for each $k \ne j$. Theorem \ref{p4ThmErProb} stated below gives an expression of the probability $\Pr\left\{ r_j > r_k \right\}$.
\begin{thm}
	\label{p4ThmErProb}
Let us assume the Rician channel
          model with $\E\{\H\}=\bar{\H}$ and that the row vectors of
          $\H$ are independent and identically distributed with
          covariance matrix $\tilde{\Sigma}_c$ and also let
          $\sigma_2^2 \ll \sigma_R^2$. When $\Pi_{\p}^i$ is
        the actual signal transmitted and $\Theta$ is a constant
        matrix independent of $\H$, the probability of the events $r_j
        > r_k$, i.e.  $\Pr\left\{ r_j > r_k \right\}$ can be
        approximated as:
	\begin{enumerate}
		\item For $k = i$:
		\begin{align}
		\Pr\left\{ r_j > r_i  \right\}  =\frac{1}{2} \left[1 - \sqrt{\frac{\beta_1}{1+\beta_1}} \sum_{n=0}^{N_R-1}\binom{2n}{n} \left( \frac{1}{4(1+\beta_1)} \right)^n\right],
		\label{p4eqn14_1}
		\end{align}

		where $\beta_1 = \frac{\left( \Pi_{\p}^i -\Pi_{\p}^j \right)^H \Theta^H\left(\tilde{\Sigma}_c + \frac{1}{N_R}\bar{\H}^H\bar{\H}\right)\Theta\left( \Pi_{\p}^i -\Pi_{\p}^j \right)}{4\sigma_R^2}$.
		\item For $k \ne i$ and when $q_R = \mathcal{R}e\{q\} \ne 0$, where $q$ is defined in (\ref{p4eqn12a_6}):

		\begin{align}
		\Pr\left\{ r_j > r_k  \right\} =   \frac{1}{2} \left[1 - \frac{1}{2} \sqrt{\frac{\beta_2}{1+\beta_2}}\sum_{n=0}^{N_R-1}\binom{2n}{n} \left( \frac{1}{4(1+\beta_2)} \right)^n \right],
		\label{p4eqn14_2}
		\end{align}

		where $\beta_2 = \frac{q_R^2}{\left(2+\sigma_{\kappa}^2\right)\sigma_{z_1}^2}$  and the constants $\sigma_{z_1}^2$ and $\sigma_{\kappa}^2$ are defined in  (\ref{p4eqn12a_4}) and (\ref{p4eqn12a_10}), respectively. 
				\item For $k \ne i$ and when $q_R =0$, $\Pr\left\{ r_j > r_k  \right\} = \frac{1}{2}$.
			\end{enumerate}
	\begin{proof}
		See Appendix \ref{p4ThmErProbProof} for proof.
	\end{proof}
\end{thm} 

Finally, for each transmitted symbol $\Pi_{\p}^i$, the minimum value of $\Pr\left\{ r_j > r_k \right\},~\forall~k \ne j$ is computed using Theorem \ref{p4ThmErProb} and it is substituted for $\Pr\left\{ \Pi_{\p}^i \rightarrow \Pi_{\p}^j \right\}$ into (\ref{p4eqn7a_1}) for achieving the bound of the average BER. Note that (\ref{p4eqn14_1}) and (\ref{p4eqn14_2}) also hold for the Rayleigh channel, in which case the matrix $\left(\tilde{\Sigma}_c + \frac{1}{N_R}\bar{\H}^H\bar{\H}\right)$ is replaced by $\I$ to calculate $\beta_1$ and $\beta_2$.
  \par 
 The bounds derived for the average BER can be used for both Scheme 1 and Scheme 2. In the case of Scheme 1 $\Omega = N_2M$ and $n_b = \log_2 M + \lfloor \log_2 N_2 \rfloor$, whereas for Scheme 2, the corresponding values are $\Omega = \binom{N_2}{N_T}M$ and $n_b = \log_2 M + \left\lfloor \log_2 \binom{N_2}{N_T} \right\rfloor$. For Scheme 3, $\Theta$ is no longer independent of $\H$ and therefore the bounds in Theorem~\ref{p4ThmErProb} do not hold. However, the conditional probabilities derived in  Appendix \ref{p4appCond} can be used for Scheme 3 also. Based on this the unconditional probabilities can be derived using sampling method for computing the bound.

\section{Simulation Results}
\label{p4sym}
Extensive simulations have been carried out to establish the performance of the proposed scheme on the system parameters. Explicitly, we studied the average BER of the proposed schemes vs. the SNR, the number of elements ($N_2$) in $IRS_2$, the number of receivers ($N_R$), the number of RSs in $IRS_1$ and the Rician factor, denoted by $K$. We have considered both the optimal ML detector and the low complexity compressed sensing detector in our performance evaluation. The system parameters used are given below. 
\par
\begin{figure}
	\centering
	\begin{tikzpicture}
	\input{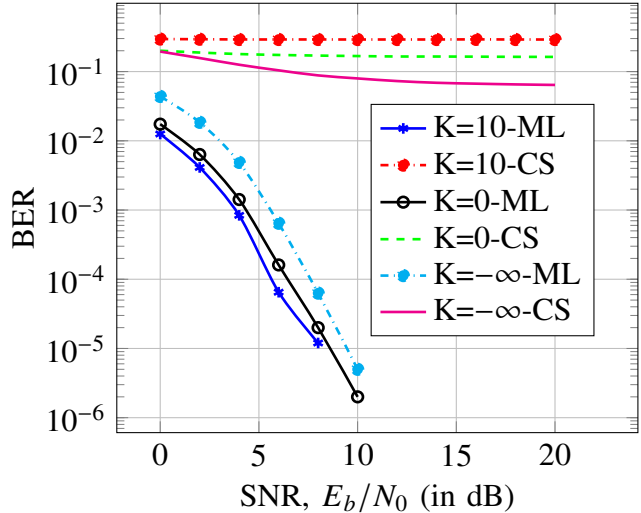}
	\end{tikzpicture}
	\caption{$S1$: Comparison of ML and CS detectors by simulations. $K$ is the Rician factor in  $dB$}
	\label{p4fig6}
\end{figure}

\begin{figure}
       	\centering
       	\begin{tikzpicture}
       	\input{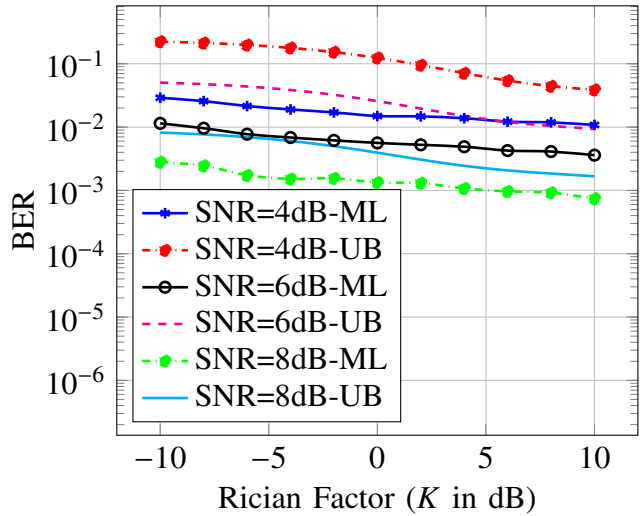}
       	\end{tikzpicture}
                   		\caption{$S1$: Comparison of the average BER and the theoretical upper bound (UB).}
       	\label{p4fig5}
       \end{figure}

\begin{figure}
                     	\centering
                     	\begin{tikzpicture}
                     	\input{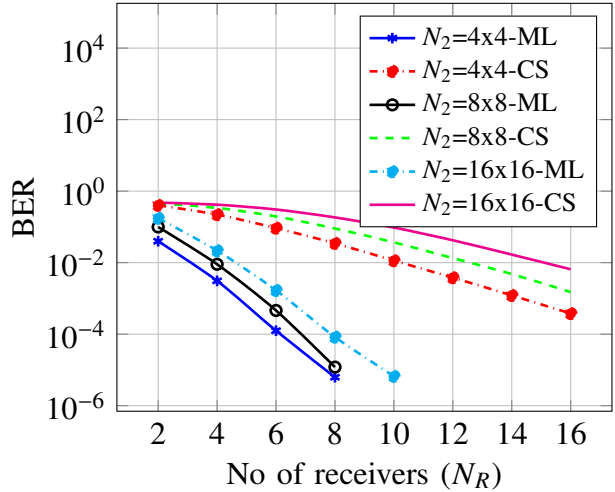}
                     	\end{tikzpicture}
                     	  	\caption{$S1$: Effect of the number of receivers ($N_R$) on the BER.}
                     	\label{p4fig7}
                     \end{figure}

\begin{figure}
                            	\centering
                            	\begin{tikzpicture}
                            	\input{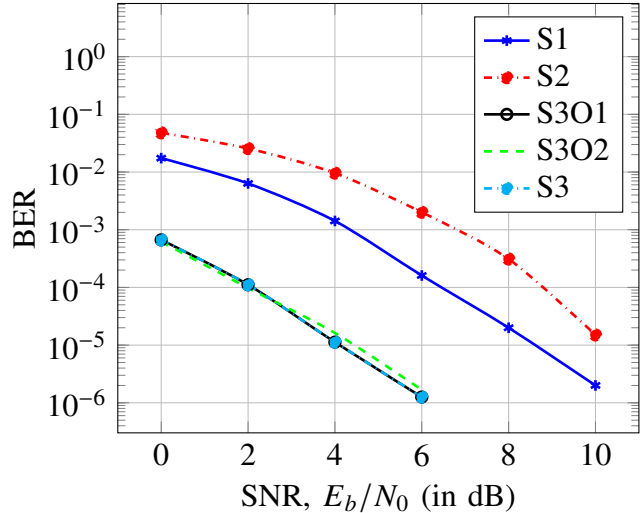}
                            	\end{tikzpicture}
                                   \caption{Comparison of three different schemes.}
                            	\label{p4fig8}
                            \end{figure}

\begin{figure}
                                   	\centering
                                   	\begin{tikzpicture}
                                   	\input{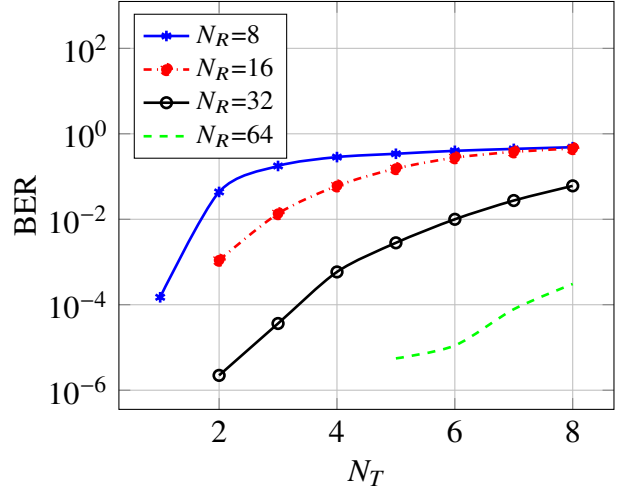}
                                   	\end{tikzpicture}
                                   	\caption{$S2$: Effect of the number of reflecting surfaces in $IRS_1$ ($N_T$) }
                                   	\label{p4fig9}
                                   \end{figure}

\begin{figure}
                               	\centering
                               	\begin{tikzpicture}
                               	\input{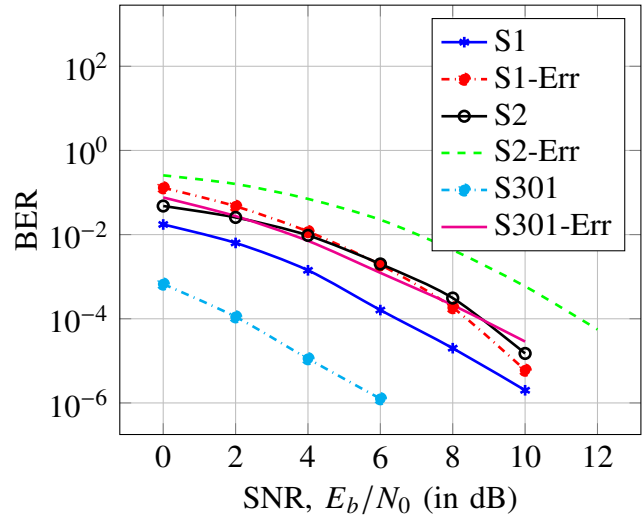}
                               	\end{tikzpicture}
                               	\caption{Effect of channel estimation error.}
                               	\label{p4fig10}
                               \end{figure}

 
\begin{itemize}
	\item $IRS_1$ : A $100 \times 100$ rectangular array with spacing $2.5~mm$. This corresponds to half wavelength of the frequency. 
	\item Distance between IRSs ($D$): $30~m$. This distance satisfies the far-field condition for $IRS_1$. 
	\item $IRS_2$: In general, an $8 \times 8$ rectangular array is used with inter element spacing of $d_2=60~cm$. However, these parameters are changed for the various performance studies, which is mentioned in the corresponding discussions. 
\end{itemize}
We used $16$ level $QAM$ at $60~GHz$  in all simulations. The results are shown in Fig.~\ref{p4fig6}-\ref{p4fig10} for $10^5$ Monte Carlo runs. We used both Rician and Rayleigh fading channels. Throughout the simulations, it is assumed that the channel is perfectly known at the receiver. Channel estimation in IRS-aided systems is quite a challenge, since the IRS is passive and has no signal processing capability. However, the schemes adopted in \cite{zhou2020framework, wang2019channel} could be used as a solution to the channel estimation problem.
In Fig.~\ref{p4fig6}-\ref{p4fig10}, the proposed schemes 1 and 2 are referred  as $S1$ and $S2$, respectively. $S3$ represents the solution of (\ref{p4eqn4}), whereas $S301$ and $S302$ represent the solution of (\ref{p4eqn4c}) and (\ref{p4eqn4f}), respectively. $S1$-Err represents the results of the proposed scheme 1 in the presence of channel estimation errors. ML and CS represent the results of optimal ML and compressed sensing based detectors, respectively and UB, the theoretical upper bound given in (\ref{p4eqn7d_1}).
\par
Fig.~\ref{p4fig6} shows the performance of
  scheme $S1$ at various SNRs for different values of the Rician factor
  ($K$). Observe from Fig.~\ref{p4fig6} that the performance of
  the low-complexity CS detector is inferior to that of the ML
  detector. As for the ML detector, the average BER tends to zero above
  $10~dB$ SNR, whereas it exhibits an error floor near $0.1$ for
  the CS detector. The performance of the CS detector is improved,
  when the Rician factor $K$ decreases. This is because as $K$
  decreases, the projection matrix $\A$ of the signal recovery
  becomes more random and the restricted isometric property
  \cite{1614066} is improved and it is best for the Rayleigh
  channel. However, the performance of the ML detector is improved
  with $K$, since the LOS component is increased with $K$. This is
  also reflected by the upper bound seen in Fig. ~\ref{p4fig5}. 
\par
Fig.~\ref{p4fig5}  shows the performance of single-symbol beam-index modulation ($S1$) against the Rician factor ($K$) for different SNRs. The  average BER obtained through simulations is compared against the upper bound derived in Section~\ref{p4ana}. The BER improvement vs. $K$ is due to increasing the LOS component. However, the variation in BER is only moderate, because the SNR is kept constant upon increasing $K$. Furthermore, the optimal ML detector has perfect channel knowledge. However, as $K$ increases, the gap between the simulation results and the corresponding upper bound is reduced.  This is also observed at high SNRs. Therefore, it can be concluded that for both these cases, our bound becomes tighter.

    \par 
The performance of the CS detector can be improved by increasing the number of RAs as shown in Fig.~\ref{p4fig7}, where the average BER is plotted against $N_R$. This is plotted for Rayleigh channel, which gives the best performance for CS detector. The curves are shown for different number of elements in $IRS_2$ (i.e. $N_2$). Observe that for the same number of RAs, the performance degrades, as $N_2$ increases.  However, as $N_2$ increases, the data rate will increase. 
        \par 
        \vspace{-0.8cm}
        Fig. ~\ref{p4fig8} compares the ML detector's performance for the three proposed schemes in terms of their average BER for Rician factor $K=0~dB$. For $S2$, we used $N_T=2$, i.e. the number of RSs in $IRS_1$ is two. This is because, if $N_T$ is large, the complexity of optimal ML decoding will escalate. For fair comparison, the number of elements in all three cases are kept the same. Therefore, for $S3$, where the optimization is to be carried out in an array, a $2 \times 2$ element array is considered to form a single RS. Hence, the effective dimension of $IRS_2$ in $S3$ is $4 \times 4$, while it is $8 \times 8$ in the case of $S1$ and $S2$.  Hence, the data rate will be lowest for $S3$, whereas it is the highest for $S2$, since there are more RSs in $IRS_1$. The data rate for $S1$, $S2$ and $S3$ are 10, 14 and 8 $bpcu$, respectively. In Fig.~\ref{p4fig8}, the legends $S3O1$ and $S3O2$ represent the results of two optimization methods, i.e. the solution of (\ref{p4eqn4c}) and that of (\ref{p4eqn4f}), respectively. Both these schemes perform better than $S1$ and $S2$. This is because there is an increase in the received SNR due to optimization. In addition in S3, the modulating symbol is embedded in $4$ elements, which gives an additional performance improvement. This makes the BER gap between the curves of $S3$ and the other schemes substantial. Observe that $S3O1$ performs marginally better than $S3O2$. This is because $S302$ maximizes the lower bound, whereas $S301$ operates on the exact function. Note that $S301$ performs almost similar to the solution of the exact equation (\ref{p4eqn4}). The performance of $S2$ is approximately $2~dB$ worse than that of $S1$. There are two differences between these two schemes. In $S2$, the same QAM/PSK symbol is carried by more than one elements in $IRS_2$. Hence, the probability of error in decoding the modulating symbol is reduced compared to $S1$. However, the information carried by the beam-index is higher in the case of $S2$, whose probability of decoding error will be higher than that of $S1$. The average BER reflects these two opposite effects. 
     \par 
     Fig.~\ref{p4fig9} shows the effect of the number of RSs ($N_T$) in $IRS_1$ on the BER performance in $S3$ for different number of receivers ($N_R$) for CS detector. Similar to Fig.~\ref{p4fig7}, this is also under Rayleigh channel condition. The average  BER increases as $N_T$ increases, which can be reduced by increasing the number of receivers. However, as $N_T$ increases, the data rate increases. In this case, for the single RS case (which is equivalent to $S1$), the data rate is $10~bpcu$, while it is  14, 23, 30 and 36 $bpcu$ for $N_T = 2, 4, 6 ~\& ~8$, respectively. Finally, in Fig. \ref{p4fig10}, the effect of channel estimation errors is demonstrated. The true channel coefficients are corrupted by adding noise having a variance of $\sigma_R^2$, which affects both the  optimization as well as detection. The average BER is shown in the figure both with and without channel estimation error. It can be seen that both $S1$ and $S2$ have approximately $2-3~dB$ performance degradation owing to the channel estimation error, whereas this gap is in excess of $4~dB$ for $S3$. This is because, in $S3$, the contaminated channel information is used both for optimization and detection.
           \par 
   
\section{Conclusions}
\label{p4concl}   
We proposed beam index modulation for millimeter wave communication exploiting the benefits of IRSs. The proposed scheme has three main advantages: 1) It achieves low-cost beamforming by using IRS for applying phase shifts, 2) it is capable of achieving reliable communication with the help of multiple IRSs in non-LOS scenarios, and 3) it sends additional information using beam-index modulation without any additional cost. Furthermore, we developed the optimal ML detector and a low-complexity compressed sensing detector for the proposed schemes. An upper bound of the average BER of the optimal ML detector is also achieved. Finally, the performance of the proposed schemes was evaluated through extensive simulations. 
\appendices
\section{Proof of Lemma \ref{p4_lem_2}}
\label{p4_lem_2_proof}
\vspace{-1cm}
  \begin{align}
  \|\H_Q\Theta_Q \b_Q s\|^2 &=  (\b_Q s)^H\Theta_Q^H\H_Q^H\H_Q\Theta_Q (\b_Q s) \nonumber \\
  &= \text{Tr} \left\{ (\b_Q s)^H\Theta_Q^H\H_Q^H\H_Q\Theta_Q (\b_Q s)\right \} = \text{Tr} \left\{ \Theta_Q^H\H_Q^H\H_Q\Theta_Q (\b_Q s) (\b_Q s)^H\right\}. 
  \label{p4eqn4d}
  \end{align}
  Note that $ \Theta_Q^H\H_Q^H\H_Q\Theta_Q$ and $(\b_Q s) (\b_Q s)^H$ are positive definite matrices. Therefore applying \cite[Theorem 2]{362841} on  (\ref{p4eqn4d}) results in (\ref{p4eqnlem2_1}). Now, since $H_Q$ is a random matrix and if $R \ge Q$, $\H_Q^H\H_Q$ will be a full-rank matrix with probability 1 and consequently $\lambda_{min}\left(\Theta_Q^H\H_Q^H\H_Q\Theta_Q \right)  > 0$. 

\section{Proof of Theorem \ref{p4ThmErProb}}
\label{p4ThmErProbProof}
Lemma \ref{p4LemQfn} of Appendix \ref{p4appCond} gives the conditional probability $\Pr\left\{ r_j > r_k |\A \right\}$. Explicitly, the conditional probabilities are in the form of complementary error function ($Q$-functions). The distributions of the arguments of these $Q$-functions are derived in Lemma \ref{p4LemConErProb} in Appendix \ref{p4Appdist}. Therefore, the unconditional probabilities can be calculated by taking expectation of conditional probabilities (\ref{p4eqn10b_1}) with respect to the corresponding distributions of their arguments in (\ref{p4eqn13_1}) and (\ref{p4eqn13_2}).  
\par 
For $k =i$, the argument of the conditional probability is a $\Gamma$-distributed random variable with parameters $\Gamma\left(N_R, \frac{2\sigma_R^2}{\left( \Pi_{\p}^i -\Pi_{\p}^j \right)^H\Theta^H\Sigma_c\Theta\left( \Pi_{\p}^i -\Pi_{\p}^j \right)}\right)$. The unconditional probability in this case is 
\begin{align}
\Pr\left\{ r_j > r_i  \right\}  = 	\int_{0}^{\infty}Q\left(\sqrt{\gamma_{ij}}\right)  \left(\frac{2\sigma_R^2}{\left( \Pi_{\p}^i -\Pi_{\p}^j \right)^H\Theta^H\Sigma_c\Theta\left( \Pi_{\p}^i -\Pi_{\p}^j \right)}\right)^{N_R}\frac{\gamma_{ij}^{N_R-1}e^{-\frac{2\sigma_R^2}{\left( \Pi_{\p}^i -\Pi_{\p}^j \right)^H\Theta^H\Sigma_c\Theta\left( \Pi_{\p}^i -\Pi_{\p}^j \right)}\gamma_{ij}}}{\Gamma\left(N_R\right)}d\gamma_{ij}.
\label{p4eqn14_3}
\end{align}
\par 
 The closed-form expression for (\ref{p4eqn14_3}) given in \cite[Eq. (A12)]{380145} can be applied to get (\ref{p4eqn14_1}). This proves the first part of the theorem. 
\par 
 For $k \ne i$, the probability is computed as follows. First we will consider the case of $q_R \ne 0$. By exploiting the relationship $Q(x) = 1- Q(-x)$, the probability $\Pr\left\{ r_j > r_k  \right\} $ can be written as:

\begin{align}
\Pr\left\{ r_j > r_k  \right\}  &= 	\int_{-\infty}^{\infty}Q\left(\kappa\right)f_{\kappa}\left(\kappa\right)d\kappa 	= 	\int_{-\infty}^{0}\left(1-Q\left(-\kappa\right)\right)f_{\kappa}\left(\kappa\right)d\kappa + 	\int_{0}^{\infty}Q\left(\kappa\right)f_{\kappa}\left(\kappa\right)d\kappa \nonumber \\
&= 	\int_{0}^{\infty}f_{\kappa}\left(-\kappa\right)d\kappa + 	\int_{0}^{\infty}Q\left(\kappa\right)\left(f_{\kappa}\left(\kappa\right)-f_{\kappa}\left(-\kappa\right)\right)d\kappa = \tilde{C} \left(I_1+I_2\right),
\label{p4eqn14_4}
\end{align}

where $\tilde{C} = \frac{2\Gamma(2N_R)}{\Gamma(N_R)\sqrt{\pi\sigma_{\kappa}^2}}\left(\frac{v-1}{2v}\right)^{N_R}$ is a constant term and 
\begin{align}
I_1 &= \int_{0}^{\infty} e^{- \frac{2v-1}{2v\sigma_{\kappa}^2}\kappa^2}D_{-2N_R}\left(\sqrt{\frac{2}{v\sigma_{\kappa}^2}}\kappa\right)d\kappa, 
\label{p4eqn14_8}
\end{align}
while 
\begin{align}
I_2 = \int_{0}^{\infty} Q(\kappa)e^{- \frac{2v-1}{2v\sigma_{\kappa}^2}\kappa^2}\left(D_{-2N_R}\left(-\sqrt{\frac{2}{v\sigma_{\kappa}^2}}\kappa\right) - D_{-2N_R}\left(\sqrt{\frac{2}{v\sigma_{\kappa}^2}}\kappa\right)\right)d\kappa.
\label{p4eqn14_9}
\end{align}
Note $D_{(.)}(.)$ is the parabolic cylinder function \cite[pp. 45]{mathai2006generalized} and it can be written in terms Kummer's confluent hypergeometric function ${}_1F_1(.)$ as \cite[pp. 39 (23)]{buchholz2013confluent}
\begin{align}
D_K(z) = 2^{\frac{K}{2}}\sqrt{\pi}e^{-\frac{z^2}{4}}\left(\frac{1}{\Gamma\left(\frac{1-K}{2}\right)} {}_1 F_1\left(-\frac{K}{2};\frac{1}{2};\frac{z^2}{2}\right)-\frac{z}{\sqrt{2}\Gamma\left(-\frac{K}{2}\right)} {}_1 F_1\left(\frac{1-K}{2};\frac{3}{2};\frac{z^2}{2}\right)\right).
\label{p4eqn14_5}
\end{align}
Let us substitute $\kappa = +\sqrt{t}$ into (\ref{p4eqn14_8}) and expand $D_{-2N_R(.)}$ using (\ref{p4eqn14_5}). Note that $d\kappa = \frac{1}{2\sqrt{t}}dt$. Therefore $I_1$ becomes:
\begin{align}
I_1 &= \frac{\sqrt{\pi}2^{-\left(N_R+1\right)}}{\Gamma\left(N_R + \frac{1}{2}\right)} \int_{0}^{\infty}t^{-\frac{1}{2}}e^{-\frac{t}{\sigma_{\kappa}^2}}~{}_1F_1\left(N_R;\frac{1}{2};\frac{t}{v\sigma_{\kappa}^2}\right)dt \nonumber \\
&- \frac{2^{-\left(N_R+1\right)}}{\Gamma\left(N_R\right)} \sqrt{\frac{\pi}{v\sigma_{\kappa}^2}}\int_{0}^{\infty} e^{-\frac{t}{\sigma_{\kappa}^2}}~{}_1F_1\left(N_R+\frac{1}{2};\frac{3}{2};\frac{t}{v\sigma_{\kappa}^2}\right) dt.
\label{p4eqn14_15}
\end{align}
Now, the difference in (\ref{p4eqn14_9}) is formulated as: 
\begin{align}
\Delta &= D_{-2N_R}\left(-\sqrt{\frac{2t}{v\sigma_{\kappa}^2}}\right) - D_{-2N_R}\left(\sqrt{\frac{2t}{v\sigma_{\kappa}^2}}\right) \nonumber \\
&=  \frac{2e^{-\frac{t}{2v\sigma_{\kappa}^2}}2^{-N_R}}{\Gamma\left(N_R\right)}\sqrt{\frac{\pi t}{v\sigma_{\kappa}^2}}~{}_1 F_1\left(N_R+\frac{1}{2};\frac{3}{2};\frac{t}{v\sigma_{\kappa}^2}\right). 
\label{p4eqn14_12}
\end{align}
	In order to evaluate $I_2$, first we express the $Q$-function in terms of the complimentary error function as $Q(x) = \frac{1}{2}\text{erfc}\left(\frac{x}{\sqrt{2}}\right)$ \cite[pp. 40]{proakis2001digital}, and then subsequently it is expressed in terms of the hypergeometric function as~\cite{wolfram1}:
	\begin{align}
	Q(x) = \frac{1}{2} - \frac{x}{\sqrt{2\pi}} ~{}_1 F_1\left(\frac{1}{2};\frac{3}{2};-\frac{x^2}{2}\right).
	\label{p4eqn14_13}
	\end{align}
	Upon substituting (\ref{p4eqn14_12}) and (\ref{p4eqn14_13}) into (\ref{p4eqn14_9}), $I_2$ becomes:
	\begin{align}
	I_2 &= \frac{2^{-\left(N_R+1\right)}}{\Gamma\left(N_R\right)} \sqrt{\frac{\pi}{v\sigma_{\kappa}^2}}\int_{0}^{\infty} e^{-\frac{t}{\sigma_{\kappa}^2}}~{}_1F_1\left(N_R+\frac{1}{2};\frac{3}{2};\frac{t}{v\sigma_{\kappa}^2}\right)dt \nonumber \\
	&- \frac{2^{-N_R}}{\Gamma\left(N_R\right)} \sqrt{\frac{1}{2v\sigma_{\kappa}^2}}\int_{0}^{\infty} t^{\frac{1}{2}}e^{-\frac{t}{\sigma_{\kappa}^2}}~{}_1 F_1\left(\frac{1}{2};\frac{3}{2};-\frac{t}{2}\right)~{}_1F_1\left(N_R+\frac{1}{2};\frac{3}{2};\frac{t}{v\sigma_{\kappa}^2}\right) dt.
	\label{p4eqn14_16}
	\end{align}
	Note that the second term of the RHS in (\ref{p4eqn14_15}) and the first term of RHS in (\ref{p4eqn14_16}) will get cancelled. Hence $\Pr\left\{ r_j > r_k  \right\} $ will become:
	\begin{align}
	\Pr\left\{ r_j > r_k  \right\}  &= C_1 \int_{0}^{\infty}t^{-\frac{1}{2}}e^{-\frac{t}{\sigma_{\kappa}^2}}~{}_1F_1\left(N_R;\frac{1}{2};\frac{t}{v\sigma_{\kappa}^2}\right)dt \nonumber \\
	&- C_2 \int_{0}^{\infty} t^{\frac{1}{2}}e^{-\frac{t}{\sigma_{\kappa}^2}}~{}_1 F_1\left(\frac{1}{2};\frac{3}{2};-\frac{t}{2}\right)~{}_1F_1\left(N_R+\frac{1}{2};\frac{3}{2};\frac{t}{v\sigma_{\kappa}^2}\right) dt \nonumber \\
	&= C_1I_3 -C_2I_4,
	\label{p4eqn15_1}
	\end{align}
	where $C_1 = \frac{1}{2\sqrt{\pi\sigma_{\kappa}^2}}\left(\frac{v-1}{v}\right)^{N_R}$ and $C_2 = \frac{1}{\pi\sigma_{\kappa}^2\sqrt{2v}}\frac{\Gamma\left(N_R+\frac{1}{2}\right)}{\Gamma(N_R)}\left(\frac{v-1}{v}\right)^{N_R}$. Note that we have exploited the relationship $\Gamma(x)\Gamma\left(x +\frac{1}{2}\right) = \frac{\sqrt{\pi}}{2^{2x-1}}\Gamma(2x)$  \cite[Theorem 6]{gacite} for reducing $C_1$ and $C_2$. Now, $I_3$ is expressed using equation \cite[pp. 822, equation(7.621(4))]{gradshteyn2014table} and it is given below:
	\begin{align}
	I_3 &= \int_{0}^{\infty}t^{-\frac{1}{2}}e^{-\frac{t}{\sigma_{\kappa}^2}}~{}_1F_1\left(N_R;\frac{1}{2};\frac{t}{v\sigma_{\kappa}^2}\right)dt \nonumber \\
	&= \Gamma\left(\frac{1}{2}\right)\sqrt{\sigma_{\kappa}^2}~{}_2F_1\left(N_R, \frac{1}{2};\frac{1}{2};\frac{1}{v} \right) = \sqrt{\pi\sigma_{\kappa}^2}\left(1 - \frac{1}{v}\right)^{-N_R},
	\label{p4eqn15_2}
	\end{align}
	where ${}_2 F_1(a,b;c;z)$ is Gauss' Hypergeometric function \cite[pp. 42]{mathai2006generalized}. Observe that ${}_2F_1\left(a, b;b;z \right) = {}_1F_0\left(a; z \right) = \left(1- z\right)^{-a}$ \cite[A1]{6389887}. In order to evaluate the integral $I_4$, we make the substitution $z = \frac{t}{v\sigma_{\kappa}^2}$ and use \cite[pp. 823, equation (7.622(1))]{gradshteyn2014table} and the integral becomes:
	\begin{align}
	I_4 &= \int_{0}^{\infty} t^{\frac{1}{2}}e^{-\frac{t}{\sigma_{\kappa}^2}}~{}_1 F_1\left(\frac{1}{2};\frac{3}{2};-\frac{t}{2}\right)~{}_1F_1\left(N_R+\frac{1}{2};\frac{3}{2};\frac{t}{v\sigma_{\kappa}^2}\right) dt \nonumber \\
	&= \left(v\sigma_{\kappa}^2\right)^{\frac{3}{2}} \int_{0}^{\infty} z^{\frac{1}{2}}e^{-vz}~{}_1 F_1\left(\frac{1}{2};\frac{3}{2};-\frac{v\sigma_{\kappa}^2}{2}z\right)~{}_1F_1\left(N_R+\frac{1}{2};\frac{3}{2};z\right) dz \nonumber \\
	&= \sqrt{\frac{\pi}{2\left(2+\sigma_{\kappa}^2\right)}}\left(\sigma_{\kappa}^2\right)^{\frac{3}{2}} \left(\frac{v-1}{v}\right)^{-\left(N_R+\frac{1}{2}\right)}~{}_2F_1\left(N_R+\frac{1}{2},\frac{1}{2};\frac{3}{2};-\frac{\sigma_{\kappa}^2}{\left(v-1\right)\left(2+\sigma_{\kappa}^2\right)}\right).
	\label{p4eqn15_3}
	\end{align}
	Now we apply the transformations  $ {}_2 F_1\left(a,b;c;z\right) = \left(1-z\right)^{-b} ~{}_2 F_1\left(c-a,b;c;\frac{z}{z-1}\right)$~\cite{wolfram2} and ${}_2 F_1\left(a,b;b+1;z\right) = bz^{-b}\mathcal{B}_z\left(b,1-a\right)$~\cite{wolfram3}, when (\ref{p4eqn15_3}) becomes:
	\begin{align}
	I_4 = \frac{\sigma_{\kappa}^2}{2}\sqrt{\frac{\pi v}{2}}\left(\frac{v-1}{v}\right)^{-N_R} \mathcal{B}_{\frac{\sigma_{\kappa}^2}{2\left(v-1\right)+v\sigma_{\kappa}^2}}\left(\frac{1}{2},N_R\right),
	\label{p4eqn15_4}
	\end{align}
where $\mathcal{B}_z(a,b) = \mathcal{B}(a,b) z^a  \sum_{k=0}^{b-1}\frac{\left(a\right)_{(k)}}{k!} \left(1 -z\right)^k$ is the incomplete Beta function \cite{wolfram4} with $ \mathcal{B}(a,b) = \frac{\Gamma(a)\Gamma(b)}{\Gamma(a+b)}$ being the Beta function and $\left(a\right)_{(k)}$ is the Pochhammer symbol. Finally upon substituting (\ref{p4eqn15_2}) and  (\ref{p4eqn15_4}) into (\ref{p4eqn15_1}), we arrive at:
\begin{align}
	\Pr\left\{ r_j > r_k  \right\}  = \frac{1}{2} - \frac{1}{4} \sqrt{\frac{\sigma_{\kappa}^2}{2\left(v-1\right)+v\sigma_{\kappa}^2}}\sum_{n=0}^{N_R-1}\frac{\left(\frac{1}{2}\right)_{(n)} }{n!} \left(1 - \frac{\sigma_{\kappa}^2}{2\left(v-1\right)+v\sigma_{\kappa}^2}\right)^n.
\label{p4eqn15_5}
\end{align}
Now substituting for $v$ and the Pochhammer symbol $\left(a\right)_{(n)} = \frac{\Gamma\left(n+a\right)}{\Gamma\left(a\right)}$ \cite{geller1969table} will give the second term in the RHS of (\ref{p4eqn14_2}). 
\par 
Finally, when $q_R =0$, the distribution of $\kappa$ is zero mean Gaussian. Hence, in this case $I_2$ in (\ref{p4eqn14_4}) will be zero, while $I_1 = \frac{1}{2}$ and the constant $\tilde{C} =1$, which completes the proof.

\section{Conditional Probability $\Pr\{r_j > r_k |\A\}$ }
\label{p4appCond}
\begin{lem}
	\label{p4LemQfn}
		Let $ \Pi_{\p}^i$ be the transmitted signal. Then, we have 
			\begin{align}
			\Pr\left\{ r_j > r_k |\A \right\}=
			\begin{cases}
			Q\left(\sqrt{\gamma_{ij}}\right), & \text{if}\ k = i \\
			Q\left(\kappa\right), & \text{otherwise}
			\end{cases},
			\label{p4eqn10b_1}
			\end{align}
			where $Q(x) = \frac{1}{\sqrt{2\pi}}\int_{x}^{\infty} \e^{-\frac{t^2}{2}}dx$ is the complementary error function and
			\begin{align}
			\gamma_{kj} =\frac{1}{2}\left( \Pi_{\p}^k -\Pi_{\p}^j \right)^H \A^H\Sigma^{-1}\A\left( \Pi_{\p}^k - \Pi_{\p}^j \right),
			\label{p4eqn10a_1}
			\end{align}
			and 
			\begin{align}
			\kappa = \frac{\gamma_{ik} - \gamma_{ij}}{\sqrt{\gamma_{kj}}}.
			\label{p4eqn10a_2}
			\end{align}
			\begin{proof}
					First the conditional probability $\Pr\left\{ r_j > r_k |\A\right\}$ is estimated for $k \ne i$ when $\Pi_{\p}^i$ is transmitted as follows.The event $r_j > r_k$ is 			
					\begin{align}
					\mathcal{R}e\left\{\left(\y - \frac{1}{2} \A \Pi_{\p}^k\right)^H\Sigma^{-1}\A \Pi_{\p}^k \right\} < \mathcal{R}e\left\{\left(\y -\frac{1}{2} \A \Pi_{\p}^j\right)^H\Sigma^{-1}\A \Pi_{\p}^j \right\}.
					\label{p4eqn8_1}
					\end{align}
					Using (\ref{p4eqn5a}), (\ref{p4eqn8_1}) can be written as:
					\begin{align}
				\mathcal{R}e\left\{\left(\w + \A \Pi_{\p}^i \right)^H \Sigma^{-1}\A\left( \Pi_{\p}^k -\Pi_{\p}^j \right) \right\} < \mathcal{R}e\left\{-\frac{1}{2}\left( \Pi_{\p}^j -\Pi_{\p}^k \right)^H \A^H\Sigma^{-1}\A\left( \Pi_{\p}^j + \Pi_{\p}^k \right)\right\}.
					\label{p4eqn9_1}
					\end{align}
					Hence, $\Pr\left\{ r_j > r_k |\A \right\} = \Pr\left\{ \eta < g_{jk} | \A \right\}$,
					where $\eta = \mathcal{R}e\left\{\left(\w + \A \Pi_{\p}^i \right)^H \Sigma^{-1}\A\left( \Pi_{\p}^k - \Pi_{\p}^j \right)\right\}$  and $g_{jk}$ is the RHS of  (\ref{p4eqn9_1}). Finally, note that $\eta \sim \N(\mu_{jk},\gamma_{jk})$, where $\mu_{jk} = \mathcal{R}e \left\{\left(\Pi_{\p}^i\right)^H \A^H\Sigma^{-1}\A\left( \Pi_{\p}^k - \Pi_{\p}^j \right)\right\}$ and $\gamma_{jk}$ is defined in (\ref{p4eqn10a_1}). Now, let us make a substitution $\tilde{\eta} = \frac{\eta -\mu_{jk}}{\sqrt{\gamma_{jk}}}$. Clearly, $\tilde{\eta}$ is a standard normal random variable and hence the probability in (\ref{p4eqn10a_1}) can be written in the form of the $Q$-function as:
					\begin{align}
					\Pr\left\{r_j > r_k | \A \right\} = \Pr\left\{\tilde{\eta}  <  -\frac{s_{jk}}{\sqrt{\gamma_{jk}}}\right\} = Q\left(\frac{s_{jk}}{\sqrt{\gamma_{jk}}}\right),
					\label{p4eqn11b_1}
					\end{align}
					where we have
						\begin{align}
						s_{jk} &= -\left(g_{jk}-\mu_{jk}\right)=\mathcal{R}e \left\{\frac{1}{2}\left( \Pi_{\p}^j -\Pi_{\p}^k \right)^H \A^H\Sigma^{-1}\A\left( \Pi_{\p}^j + \Pi_{\p}^k -2\Pi_{\p}^i \right)\right\} \nonumber \\
						&= \mathcal{R}e \left\{\frac{1}{2}\left( \left(\Pi_{\p}^j - \Pi_{\p}^i \right)-\left(\Pi_{\p}^k -\Pi_{\p}^i \right) \right)^H \A^H\Sigma^{-1}\A \left(\left(\Pi_{\p}^j - \Pi_{\p}^i \right)+\left(\Pi_{\p}^k -\Pi_{\p}^i \right) \right)\right\} \nonumber \\
						&= \gamma_{ij} - \gamma_{ik}.
						\label{p4eqn11c_1}
						\end{align}

					\par 
					When $k=i$, we have $\gamma_{ik}=0$ and therefore (\ref{p4eqn11b_1}) becomes $\Pr\left\{r_j > r_i | \A \right\} = Q\left(\sqrt{\gamma_{ij}}\right).$
					This completes the proof. 
			\end{proof}
\end{lem}
\section{Distribution of $\gamma_{mn}$ and $\kappa$}
\label{p4Appdist}
\begin{lem}
	\label{p4LemConErProb}
	Under the assumptions of Theorem \ref{p4ThmErProb},
	\begin{enumerate}
		\item  The random variable $\gamma_{kj}$ defined in (\ref{p4eqn10a_1}) is a $\Gamma$-distributed random variable, i.e., 
		\begin{align}
		\gamma_{kj} \sim \Gamma\left(N_R, \frac{2\sigma_R^2}{\left( \Pi_{\p}^k -\Pi_{\p}^j \right)^H\Theta^H\Sigma_c\Theta\left( \Pi_{\p}^k -\Pi_{\p}^j \right)}\right).
		\label{p4eqn13_1}
		\end{align}
		\item When $q_R \ne 0$, the distribution of the random variable $\kappa$ defined in (\ref{p4eqn10a_2}) is 
			\begin{align}
			f_{\kappa}\left(\kappa\right) =\frac{2\Gamma(2N_R)e^{-\frac{2v-1}{2v\sigma_{\kappa}^2}\kappa^2}}{\Gamma(N_R)\sqrt{\pi\sigma_{\kappa}^2}}\left(\frac{v-1}{2v}\right)^{N_R}D_{-2N_R}\left(-\sqrt{\frac{2}{v\sigma_{\kappa}^2}}\kappa\right),
			\label{p4eqn13_2}
			\end{align}
		where  $D_{K}(.)$ is the Parabolic cylinder function \cite[pp. 45]{mathai2006generalized}  and the parameters are defined in Theorem \ref{p4ThmErProb}.  
		\item For $q_R =0$, $\kappa \sim \N(0,\frac{\sigma_{\kappa}^2}{2})$, where $\sigma_{\kappa}^2$ is defined in (\ref{p4eqn12a_10})
		 
	\end{enumerate}
	\begin{proof}
			When we have $\sigma_2^2 \ll \sigma_R^2$, $\gamma_{kj}$ in (\ref{p4eqn10a_1}) can be approximated as:
			\begin{align}
			\gamma_{kj} \approx \frac{1}{2\sigma_R^2}\left( \Pi_{\p}^k -\Pi_{\p}^j \right)^H \A^H\A\left( \Pi_{\p}^k - \Pi_{\p}^j \right).
			\label{p4eqn12a_1}
			\end{align}
		Recall that $\A = \H\Theta$. Under the assumptions of Theorem \ref{p4ThmErProb}, $\H^H\H$ obeys a non-central complex Wishart distribution, i.e.,  $\H^H\H \sim \mathcal{W}\left(N_R, \bar{\H},\tilde{\Sigma}_c\right)$. This can be approximated as a central complex Wishart distribution having the covariance matrix of $\Sigma_c = \tilde{\Sigma}_c + \frac{1}{N_R}\bar{\H}^H\bar{\H}$ \cite{xu2009approximation}\footnote{For Rayleigh channel, $\H^H\H$ exactly follows the central complex Wishart distribution associated with $\Sigma_c = \I$}. If we use this approximation, it can be assumed that  each row of $\H$ is distributed according to $\CN(0,\Sigma_c)$. Hence, the random variable $\z = \frac{1}{\sqrt{2\sigma_R^2}}\A\left( \Pi_{\p}^i -\Pi_{\p}^j \right)$ is distributed according to $\CN\left(0,\frac{1}{2\sigma_R^2}\left( \Pi_{\p}^i -\Pi_{\p}^j \right)^H\Theta^H\Sigma_c\Theta\left( \Pi_{\p}^i -\Pi_{\p}^j \right)\I \right)$.  Hence, $\gamma = \z^H\z$ is a Gamma distributed variable having the distribution function of (\ref{p4eqn13_1}). This proves the first part of the Lemma.
			\par 
			The  rest of the Lemma is proved as follows. Define $\z_1 = \frac{1}{\sqrt{2\sigma_R^2}}\A\left( \left(\Pi_{\p}^j - \Pi_{\p}^i \right)-\left(\Pi_{\p}^k -\Pi_{\p}^i \right) \right)$ and $\z_2 = \frac{1}{\sqrt{2\sigma_R^2}}\A\left( \left(\Pi_{\p}^j - \Pi_{\p}^i \right)+\left(\Pi_{\p}^k -\Pi_{\p}^i \right) \right)$. Similar to the case of first part of this Lemma, the distributions of $\z_1$ and $\z_2$ can be approximated as $\z_1 \sim \CN(0,\sigma_{z_1}^2 \I)$ and $\z_2 \sim \CN(0,\sigma_{z_2}^2\I)$, where 
			\begin{align}
				\sigma_{z_1}^2 =  \frac{\left(\Pi_{\p}^j - \Pi_{\p}^k \right)^H\Theta^H\Sigma_c\Theta\left(\Pi_{\p}^j - \Pi_{\p}^k \right)}{2\sigma_R^2},
			\label{p4eqn12a_4}
			\end{align}
			and 
			\begin{align}
			\sigma_{z_2}^2 =\frac{\left(\Pi_{\p}^j + \Pi_{\p}^k -2\Pi_{\p}^i\right)^H\Theta^H\Sigma_c\Theta\left(\Pi_{\p}^j + \Pi_{\p}^k -2\Pi_{\p}^i\right)}{2\sigma_R^2}.
			\label{p4eqn12a_5}
			\end{align}
			Also note that $\E\left\{ \z_1\z_2^H \right\}  = q\I$ and $\E\left\{ \z_2\z_1^H \right\}  = q^H\I$, where
			\begin{small}
				\begin{align}
				q = \frac{1}{2\sigma_R^2}Tr\left\{ \left(\Pi_{\p}^j - \Pi_{\p}^k \right)\left( \Pi_{\p}^j + \Pi_{\p}^k  -2\Pi_{\p}^i \right)^H \Theta^H\Sigma_c\Theta\right\} = \frac{ \left( \Pi_{\p}^j + \Pi_{\p}^k  -2\Pi_{\p}^i \right) ^H\Theta^H\Sigma_c\Theta \left(\Pi_{\p}^j - \Pi_{\p}^k \right)}{2\sigma_R^2}.
				\label{p4eqn12a_6}
				\end{align}
			\end{small}
			Now we can rewrite $\gamma_{kj}$ in (\ref{p4eqn12a_1}) as:
			\begin{align}
			\gamma_{kj} \approx \frac{1}{2\sigma_R^2}\left( \left(\Pi_{\p}^j - \Pi_{\p}^i\right)-\left(\Pi_{\p}^k - \Pi_{\p}^i \right) \right)^H \A^H \A \left( \left(\Pi_{\p}^j - \Pi_{\p}^i\right)-\left(\Pi_{\p}^k - \Pi_{\p}^i \right) \right) = \|\z_1\|^2.
			\label{p4eqn12a_7}
			\end{align}
		Similarly $\kappa \approx \mathcal{R}e\left\{\frac{\z_1^H\z_2}{\|\z_1\|}\right\}$.
		First we derive the distribution of $\kappa$ given $\z_1$. Note that $\z_1$ and $\z_2$ are complex Gaussian distributed random vectors. Hence, the distribution $f_{\kappa|\z_1}(\kappa|\z_1)$ is Gaussian. Let $\tilde{\kappa} = \frac{\z_1^H\z_2}{\|\z_1\|}$. Therefore, using \cite[prop. 3.13]{eaton1983multivariate}, we have $\mu_{\tilde{\kappa} } = \E\{ \tilde{\kappa} |\z_1 = \tilde{\z}\} =\frac{q^H}{\sigma_{z_1}^2}\|\tilde{\z}\|.$
		Note that $\kappa = \mathcal{R}e\left\{\tilde{\kappa} \right\}$ and hence $\E\{ \kappa|\z_1 = \tilde{\z}\} =  \mathcal{R}e\left\{\mu_{\tilde{\kappa} }\right\} = \frac{q_R}{\sigma_{z_1}^2}\|\tilde{\z}\|$, where $q_R=\mathcal{R}e\{q\}$. In order to compute variance of the $\kappa$ given $\z_1$, let us expand $\kappa$ as:
		\begin{align}
		\kappa =  \mathcal{R}e\left\{\frac{\z_1^H\z_2}{\|\z_1\|}\right\} = \frac{\mathcal{R}e\{\z_1\}^T	\mathcal{R}e\{\z_2\}}{\|\z_1\|} +  \frac{\mathcal{I}m\{\z_1\}^T\mathcal{I}m\{\z_2\}}{\|\z_1\|} = u_1 + u_2.
		\label{p4eqn12a_10_1}
		\end{align}
		Hence,
		\begin{align}
		\text{var}\left(\kappa |\z_1 = \tilde{\z}\right) = \text{var}\left(u_1|\z_1 = \tilde{\z}\right)+\text{var}\left(u_2|\z_1 = \tilde{\z}\right) + \text{Cov}\left(u_1,u_2|\z_1 = \tilde{\z}\right)+\text{Cov}\left(u_2,u_1|\z_1 = \tilde{\z}\right).
		\label{p4eqn12a_10_2}
		\end{align}
		Using \cite[prop. 3.13]{eaton1983multivariate}, it can be shown that: 
		\begin{align}
		\text{var}\left(u_1|\z_1 = \tilde{\z}\right) =\frac{\mathcal{R}e\{\tilde{\z}\}^T\text{Cov} \{\mathcal{R}e\{\z_2\} |\z_1= \tilde{\z}\} \mathcal{R}e\{\tilde{\z}\}}{\|\tilde{\z}\|^2}=  \frac{\mathcal{R}e\{\tilde{\z}\}^2}{\|\tilde{\z}\|^2}\frac{1}{2} \left(\sigma_{z_2}^2  - \frac{\|q\|^2}{2\sigma_{z_1}^2}\right).
		\label{p4eqn12a_10_4}
		\end{align}
		Similarly, $\text{var}\left(u_2|\z_1 = \tilde{\z}\right) =  \frac{\mathcal{I}m\{\tilde{\z}\}^2}{\|\tilde{\z}\|^2}\frac{1}{2} \left(\sigma_{z_2}^2  - \frac{\|q\|^2}{2\sigma_{z_1}^2}\right)$ and 
		\begin{align}
		\text{Cov}\left(u_1,u_2|\z_1 = \tilde{\z}\right) =	-\text{Cov}\left(u_2,u_1|\z_1 = \tilde{\z}\right) =-j\frac{\|q\|^2}{4\sigma_{z_1}^2}\frac{\mathcal{R}e\{\z_1\}^T\mathcal{I}m\{\z_1\}}{\|\z_1\|^2}. 
		\label{p4eqn12a_10_6}
		\end{align}
		Therefore, $\text{var}\left(\kappa |\z_1 = \tilde{\z}\right) = \frac{1}{2}\left( \sigma_{z_2}^2 - \frac{\|q\|^2}{2\sigma_{z_1}^2}\right)$. Hence, $\kappa|\z_1 \sim \N\left(\frac{q_R}{\sigma_{z_1}^2}\|\tilde{\z}\|,\frac{\sigma_{\kappa}^2}{2}\right)$, where
		\begin{align}
		\sigma_{\kappa}^2 = \left( \sigma_{z_2}^2 - \frac{\|q\|^2}{2\sigma_{z_1}^2}\right).
		\label{p4eqn12a_10}
		\end{align}
		If $q_R =0$, $f_{\kappa|\z_1}$ is independent of $\z_1$ and hence the unconditional distribution of $\kappa$ is the same as the conditional distribution, i.e., $\kappa \sim \N(0,\frac{\sigma_{\kappa}^2}{2})$.
			\par 
			
			When $q_R \ne 0$,  the unconditional distribution can be obtained by eliminating the conditioning with respect to the distribution of $\|\tilde{z}\|$, which is Nakagami distributed $Nakagami\left(N_R, N_R\sigma_{z_1}^2\right)$, since  $\|\tilde{z}\|^2 \sim \Gamma(N_R,\frac{1}{\sigma_{z_1}^2})$. Explicitly, $f_{\bar{Z}}\left(\bar{z}\right) = \frac{2\bar{z}^{2N_R-1}e^{-\frac{\bar{z}^2}{\sigma_{z_1}^2}}}{\Gamma(N_R)\left(\sigma_{z_1}^2\right)^{N_R}}$ \cite{nakagami1960m}, where $\bar{z} = \|\tilde{z}\|$. Hence, the unconditional distribution of $\kappa$ is:
			\begin{align}
			f_{\kappa}(\kappa) = \frac{2\int_{0}^{\infty} e^{-g(\kappa,\bar{z}) } \bar{z}^{2N_R-1}d\bar{z}}{\left(\sigma_{z_1}^2\right)^{N_R}\Gamma(N_R)\sqrt{(\pi\sigma_{\kappa}^2)}} ,
			\label{p4eqn12a_11}
			\end{align}
			where 
			\begin{align}
			g(\kappa,\bar{z}) = \frac{\left(\kappa -\frac{q_R}{\sigma_{z_1}^2}\bar{z}\right)^2}{\sigma_{\kappa}^2} +\frac{\bar{z}^2}{\sigma_{z_1}^2} = \frac{\kappa^2}{\sigma_{\kappa}^2}+ \frac{q_R^2+\sigma_{z_1}^2\sigma_{\kappa}^2}{\left(\sigma_{z_1}^2\right)^2\sigma_{\kappa}^2}\bar{z}^2-\frac{2q_R\kappa}{\sigma_{z_1}^2\sigma_{\kappa}^2}\bar{z}.
			\label{p4eqn12a_12}
			\end{align}
			Now using \cite[3.462(1)]{gradshteyn2014table}, we arrive at: 
	
			\begin{align}
				I_{\kappa} &= \int_{0}^{\infty}e^{-g(\kappa,\bar{z}) } \bar{z}^{2N_R-1}d\bar{z} =e^{-\frac{\kappa^2}{\sigma_{\kappa}^2}} \int_{0}^{\infty}e^{-\frac{q_R^2+\sigma_{z_1}^2\sigma_{\kappa}^2}{\left(\sigma_{z_1}^2\right)^2\sigma_{\kappa}^2}\bar{z}^2+\frac{2q_R\kappa}{\sigma_{z_1}^2\sigma_{\kappa}^2}\bar{z}}\bar{z}^{2N_R-1}d\bar{z}\nonumber \\
				&= e^{-\frac{\kappa^2}{\sigma_{\kappa}^2}}\left(\frac{\sigma_{z_1}^2(v-1)}{2v}\right)^{N_R}\Gamma(2N_R)e^{\frac{\kappa^2}{2\sigma_{\kappa}^2v}}D_{-2N_R}\left(-\sqrt{\frac{2}{\sigma_{\kappa}^2v}}\kappa\right), 
			\label{p4eqn12a_13}
			\end{align}
		
	where $v = 1+ \frac{\sigma_{\kappa}^2\sigma_{z_1}^2}{q_R^2}$. Finally, substituting (\ref{p4eqn12a_13}) into (\ref{p4eqn12a_11}) will give (\ref{p4eqn13_2}). 
	\end{proof}
\end{lem}

      \bibliography{lis} 

\begin{thebibliography}{10}

\bibitem{tariq2019speculative}
F.~Tariq, M.~Khandaker, K.-K. Wong, M.~Imran, M.~Bennis, and M.~Debbah, ``{A
  speculative study on 6G},'' {\em arXiv preprint arXiv:1902.06700}, 2019.

\bibitem{6732923}
S.~{Rangan}, T.~S. {Rappaport}, and E.~{Erkip}, ``Millimeter-wave cellular
  wireless networks: Potentials and challenges,'' {\em Proceedings of the
  IEEE}, vol.~102, pp.~366--385, March 2014.

\bibitem{6515173}
T.~S. {Rappaport}, S.~{Sun}, R.~{Mayzus}, H.~{Zhao}, Y.~{Azar}, K.~{Wang},
  G.~N. {Wong}, J.~K. {Schulz}, M.~{Samimi}, and F.~{Gutierrez}, ``{Millimeter
  Wave Mobile Communications for {5G} Cellular: It Will Work!},'' {\em IEEE
  Access}, vol.~1, pp.~335--349, 2013.

\bibitem{rappaport2016spectrum}
T.~S. Rappaport, ``Spectrum frontiers: The new world of millimeter-wave mobile
  communication,'' {\em Invited keynote presentation, The Federal
  Communications Commission (FCC) Headquarters}, vol.~10, 2016.

\bibitem{6392842}
``{Part 11: Wireless LAN medium access control (MAC) and physical layer (PHY)
  specifications amendment 3: enhancements for very high throughput in the 60
  GHz band},'' {\em {IEEE Std 802.11ad-2012 (Amendment to IEEE Std 802.11-2012,
  as amended by IEEE Std 802.11ae-2012 and IEEE Std 802.11aa-2012)}},
  pp.~1--628, Dec 2012.

\bibitem{6736750}
W.~{Roh}, J.~{Seol}, J.~{Park}, B.~{Lee}, J.~{Lee}, Y.~{Kim}, J.~{Cho},
  K.~{Cheun}, and F.~{Aryanfar}, ``Millimeter-wave beamforming as an enabling
  technology for {5G} cellular communications: theoretical feasibility and
  prototype results,'' {\em IEEE Communications Magazine}, vol.~52,
  pp.~106--113, February 2014.

\bibitem{7999294}
T.~S. {Rappaport}, Y.~{Xing}, G.~R. {MacCartney}, A.~F. {Molisch},
  E.~{Mellios}, and J.~{Zhang}, ``Overview of millimeter wave communications
  for fifth-generation ({5G}) wireless networks-with a focus on propagation
  models,'' {\em IEEE Transactions on Antennas and Propagation}, vol.~65,
  pp.~6213--6230, Dec 2017.

\bibitem{7522613}
A.~I. {Sulyman}, A.~{Alwarafy}, G.~R. {MacCartney}, T.~S. {Rappaport}, and
  A.~{Alsanie}, ``Directional radio propagation path loss models for
  millimeter-wave wireless networks in the {28-, 60-, and 73-GHz} bands,'' {\em
  IEEE Transactions on Wireless Communications}, vol.~15, pp.~6939--6947, Oct
  2016.

\bibitem{dai2019reconfigurable}
L.~Dai, B.~Wang, M.~Wang, X.~Yang, J.~Tan, S.~Bi, S.~Xu, F.~Yang, Z.~Chen,
  M.~Di~Renzo, {\em et~al.}, ``Reconfigurable intelligent surface-based
  wireless communication: Antenna design, prototyping and experimental
  results,'' {\em arXiv preprint arXiv:1912.03620}, 2019.

\bibitem{qingqing2019towards}
W.~Qingqing and Z.~Rui, ``Towards smart and reconfigurable environment:
  Intelligent reflecting surface aided wireless network,'' {\em arXiv preprint
  arXiv:1905.00152}, 2019.

\bibitem{di2019smart}
M.~Di~Renzo, M.~Debbah, D.-T. Phan-Huy, A.~Zappone, M.-S. Alouini, C.~Yuen,
  V.~Sciancalepore, G.~C. Alexandropoulos, J.~Hoydis, H.~Gacanin, {\em et~al.},
  ``{Smart radio environments empowered by reconfigurable AI meta-surfaces: an
  idea whose time has come},'' {\em EURASIP Journal on Wireless Communications
  and Networking}, vol.~2019, no.~1, pp.~1--20, 2019.

\bibitem{huang2019holographic}
C.~Huang, S.~Hu, G.~C. Alexandropoulos, A.~Zappone, C.~Yuen, R.~Zhang,
  M.~Di~Renzo, and M.~Debbah, ``{Holographic MIMO surfaces for 6G wireless
  networks: Opportunities, challenges, and trends},'' {\em arXiv preprint
  arXiv:1911.12296}, 2019.

\bibitem{6206517}
L.~{Subrt} and P.~{Pechac}, ``Controlling propagation environments using
  intelligent walls,'' in {\em 2012 6th European Conference on Antennas and
  Propagation (EUCAP)}, pp.~1--5, March 2012.

\bibitem{7510962}
X.~{Tan}, Z.~{Sun}, J.~M. {Jornet}, and D.~{Pados}, ``Increasing indoor
  spectrum sharing capacity using smart reflect-array,'' in {\em 2016 IEEE
  International Conference on Communications (ICC)}, pp.~1--6, May 2016.

\bibitem{yan2019passive}
W.~Yan, X.~Kuai, X.~Yuan, {\em et~al.}, ``Passive beamforming and information
  transfer via large intelligent surface,'' {\em arXiv preprint
  arXiv:1905.01491}, 2019.

\bibitem{yu2019miso}
X.~{Yu}, D.~{Xu}, and R.~{Schober}, ``{MISO} wireless communication systems via
  intelligent reflecting surfaces : ({Invited} paper),'' in {\em 2019 IEEE/CIC
  International Conference on Communications in China (ICCC)}, pp.~735--740,
  Aug 2019.

\bibitem{han2019large}
Y.~{Han}, W.~{Tang}, S.~{Jin}, C.~{Wen}, and X.~{Ma}, ``Large intelligent
  surface-assisted wireless communication exploiting statistical {CSI},'' {\em
  IEEE Transactions on Vehicular Technology}, vol.~68, pp.~8238--8242, Aug
  2019.

\bibitem{huang2019reconfigurable}
C.~{Huang}, A.~{Zappone}, G.~C. {Alexandropoulos}, M.~{Debbah}, and C.~{Yuen},
  ``Reconfigurable intelligent surfaces for energy efficiency in wireless
  communication,'' {\em IEEE Transactions on Wireless Communications}, vol.~18,
  pp.~4157--4170, Aug 2019.

\bibitem{huang2018energy}
C.~Huang, G.~C. Alexandropoulos, A.~Zappone, M.~Debbah, and C.~Yuen, ``{Energy
  efficient multi-user MISO communication using low resolution large
  intelligent surfaces},'' in {\em 2018 IEEE Globecom Workshops (GC Wkshps)},
  pp.~1--6, IEEE, 2018.

\bibitem{8811733}
Q.~{Wu} and R.~{Zhang}, ``Intelligent reflecting surface enhanced wireless
  network via joint active and passive beamforming,'' {\em IEEE Transactions on
  Wireless Communications}, vol.~18, pp.~5394--5409, Nov 2019.

\bibitem{chen2019intelligent}
J.~{Chen}, Y.~{Liang}, Y.~{Pei}, and H.~{Guo}, ``Intelligent reflecting
  surface: A programmable wireless environment for physical layer security,''
  {\em IEEE Access}, vol.~7, pp.~82599--82612, 2019.

\bibitem{pan2019multicell}
C.~Pan, H.~Ren, K.~Wang, W.~Xu, M.~Elkashlan, A.~Nallanathan, and L.~Hanzo,
  ``Multicell {MIMO} communications relying on intelligent reflecting
  surface,'' {\em Online] https://arxiv. org/abs/1907.10864}, 2019.

\bibitem{pan2019power}
C.~Pan, H.~Ren, K.~Wang, M.~Elkashlan, A.~Nallanathan, J.~Wang, and L.~Hanzo,
  ``Intelligent reflecting surface enhanced {MIMO} broadcasting for
  simultaneous wireless information and power transfer,'' {\em arXiv preprint
  arXiv:1908.04863}, 2019.

\bibitem{jiang2019over}
T.~Jiang and Y.~Shi, ``Over-the-air computation via intelligent reflecting
  surfaces,'' {\em arXiv preprint arXiv:1904.12475}, 2019.

\bibitem{bai2019latency}
T.~Bai, C.~Pan, Y.~Deng, M.~Elkashlan, A.~Nallanathan, and L.~Hanzo, ``Latency
  minimization for intelligent reflecting surface aided mobile edge
  computing,'' {\em arXiv preprint arXiv:1910.07990}, 2019.

\bibitem{basar2020reconfigurable}
E.~Basar, ``{Reconfigurable intelligent surface-based index modulation: A new
  beyond MIMO paradigm for 6G},'' {\em IEEE Transactions on Communications},
  vol.~68, no.~5, pp.~3187--3196, 2020.

\bibitem{8325484}
Y.~{Ding}, V.~{Fusco}, A.~{Shitvov}, Y.~{Xiao}, and H.~{Li}, ``Beam index
  modulation wireless communication with analog beamforming,'' {\em IEEE
  Transactions on Vehicular Technology}, vol.~67, pp.~6340--6354, July 2018.

\bibitem{4382913}
R.~Y. {Mesleh}, H.~{Haas}, S.~{Sinanovic}, C.~W. {Ahn}, and S.~{Yun}, ``Spatial
  modulation,'' {\em IEEE Transactions on Vehicular Technology}, vol.~57,
  pp.~2228--2241, July 2008.

\bibitem{6823072}
P.~{Yang}, M.~{Di Renzo}, Y.~{Xiao}, S.~{Li}, and L.~{Hanzo}, ``Design
  guidelines for spatial modulation,'' {\em IEEE Communications Surveys
  Tutorials}, vol.~17, pp.~6--26, Firstquarter 2015.

\bibitem{7342886}
S.~{Kutty} and D.~{Sen}, ``Beamforming for millimeter wave communications: An
  inclusive survey,'' {\em IEEE Communications Surveys Tutorials}, vol.~18,
  pp.~949--973, Secondquarter 2016.

\bibitem{6955833}
Z.~{Xiao}, X.~{Xia}, D.~{Jin}, and N.~{Ge}, ``Iterative eigenvalue
  decomposition and multipath-grouping {Tx/Rx} joint beamformings for
  millimeter-wave communications,'' {\em IEEE Transactions on Wireless
  Communications}, vol.~14, pp.~1595--1607, March 2015.

\bibitem{6957140}
O.~{Jo}, W.~{Hong}, S.~T. {Choi}, S.~{Chang}, C.~{Kweon}, J.~{Oh}, and
  K.~{Cheun}, ``Holistic design considerations for environmentally adaptive 60
  {GHz} beamforming technology,'' {\em IEEE Communications Magazine}, vol.~52,
  pp.~30--38, Nov 2014.

\bibitem{8468205}
Y.~{Li}, J.~{Luo}, M.~H. {Castañeda Garcia}, R.~{Böhnke}, R.~A.
  {Stirling-Gallacher}, W.~{Xu}, and G.~{Caire}, ``On the beamformed
  broadcasting for millimeter wave cell discovery: Performance analysis and
  design insight,'' {\em IEEE Transactions on Wireless Communications},
  vol.~17, pp.~7620--7634, Nov 2018.

\bibitem{7390101}
Z.~{Xiao}, T.~{He}, P.~{Xia}, and X.~{Xia}, ``Hierarchical codebook design for
  beamforming training in millimeter-wave communication,'' {\em IEEE
  Transactions on Wireless Communications}, vol.~15, pp.~3380--3392, May 2016.

\bibitem{8962355}
I.~{Aykin} and M.~{Krunz}, ``Efficient beam sweeping algorithms and initial
  access protocols for millimeter-wave networks,'' {\em IEEE Transactions on
  Wireless Communications}, pp.~1--1, 2020.

\bibitem{8978709}
Z.~{Sha}, Z.~{Wang}, S.~{Chen}, and L.~{Hanzo}, ``Graph theory based beam
  scheduling for inter-cell interference avoidance in mmwave cellular
  networks,'' {\em IEEE Transactions on Vehicular Technology}, pp.~1--1, 2020.

\bibitem{8283647}
Y.~{Ju}, H.~{Wang}, T.~{Zheng}, Q.~{Yin}, and M.~H. {Lee}, ``Safeguarding
  millimeter wave communications against randomly located eavesdroppers,'' {\em
  IEEE Transactions on Wireless Communications}, vol.~17, pp.~2675--2689, April
  2018.

\bibitem{7870294}
B.~{Sadhu}, Y.~{Tousi}, J.~{Hallin}, S.~{Sahl}, S.~{Reynolds}, O.~{Renstrom},
  K.~{Sjogren}, O.~{Haapalahti}, N.~{Mazor}, B.~{Bokinge}, G.~{Weibull},
  H.~{Bengtsson}, A.~{Carlinger}, E.~{Westesson}, J.~{Thillberg}, L.~{Rexberg},
  M.~{Yeck}, X.~{Gu}, D.~{Friedman}, and A.~{Valdes-Garcia}, ``{7.2 A 28GHz
  32-element phased-array transceiver IC with concurrent dual polarized beams
  and 1.4 degree beam-steering resolution for 5G communication},'' in {\em 2017
  IEEE International Solid-State Circuits Conference (ISSCC)}, pp.~128--129,
  Feb 2017.

\bibitem{8690621}
C.~{Scarborough}, K.~{Venugopal}, A.~{Alkhateeb}, and R.~W. {Heath},
  ``Beamforming in millimeter wave systems: Prototyping and measurement
  results,'' in {\em 2018 IEEE 88th Vehicular Technology Conference
  (VTC-Fall)}, pp.~1--5, Aug 2018.

\bibitem{7010533}
S.~{Han}, C.~{I}, Z.~{Xu}, and C.~{Rowell}, ``Large-scale antenna systems with
  hybrid analog and digital beamforming for millimeter wave {5G},'' {\em IEEE
  Communications Magazine}, vol.~53, pp.~186--194, January 2015.

\bibitem{8901444}
J.~{Zhang}, Y.~{Huang}, J.~{Wang}, R.~{Schober}, and L.~{Yang},
  ``Power-efficient beam designs for millimeter wave communication systems,''
  {\em IEEE Transactions on Wireless Communications}, vol.~19, pp.~1265--1279,
  Feb 2020.

\bibitem{7491314}
R.~{Rajashekar} and L.~{Hanzo}, ``Hybrid beamforming in mm-wave {MIMO} systems
  having a finite input alphabet,'' {\em IEEE Transactions on Communications},
  vol.~64, pp.~3337--3349, Aug 2016.

\bibitem{8924932}
A.~M. {Elbir} and K.~V. {Mishra}, ``Joint antenna selection and hybrid
  beamformer design using unquantized and quantized deep learning networks,''
  {\em IEEE Transactions on Wireless Communications}, pp.~1--1, 2019.

\bibitem{8959381}
C.~{Zhao}, Y.~{Cai}, A.~{Liu}, M.~{Zhao}, and L.~{Hanzo}, ``Mobile edge
  computing meets mmwave communications: Joint beamforming and resource
  allocation for system delay minimization,'' {\em IEEE Transactions on
  Wireless Communications}, pp.~1--1, 2020.

\bibitem{8643353}
K.~{Satyanarayana}, M.~{El-Hajjar}, A.~A.~M. {Mourad}, and L.~{Hanzo},
  ``Multi-user hybrid beamforming relying on learning-aided link-adaptation for
  mmwave systems,'' {\em IEEE Access}, vol.~7, pp.~23197--23209, 2019.

\bibitem{8964330}
K.~{Ying}, Z.~{Gao}, S.~{Lyu}, Y.~{Wu}, H.~{Wang}, and M.~{Alouini},
  ``{GMD}-based hybrid beamforming for large reconfigurable intelligent surface
  assisted millimeter-wave massive {MIMO},'' {\em IEEE Access}, vol.~8,
  pp.~19530--19539, 2020.

\bibitem{8883297}
S.~{Dutta}, C.~N. {Barati}, D.~{Ramirez}, A.~{Dhananjay}, J.~F. {Buckwalter},
  and S.~{Rangan}, ``A case for digital beamforming at mmwave,'' {\em IEEE
  Transactions on Wireless Communications}, vol.~19, pp.~756--770, Feb 2020.

\bibitem{canbilen2020reconfigurable}
A.~E. Canbilen, E.~Basar, and S.~S. Ikki, ``Reconfigurable intelligent
  surface-assisted space shift keying,'' {\em IEEE Wireless Communications
  Letters}, 2020.

\bibitem{yang2020mimo}
X.~Yang, C.-K. Wen, and S.~Jin, ``{MIMO detection for reconfigurable
  intelligent surface-assisted millimeter wave systems},'' {\em arXiv preprint
  arXiv:2004.06001}, 2020.

\bibitem{sethi2013millimeter}
W.~T. Sethi, H.~Vettikalladi, and M.~A. Alkanhal, ``{Millimeter wave antenna
  with mounted horn integrated on FR4 for 60 GHz Gbps communication systems},''
  {\em International Journal of Antennas and Propagation}, vol.~2013, 2013.

\bibitem{zhou2020framework}
G.~Zhou, C.~Pan, H.~Ren, K.~Wang, and A.~Nallanathan, ``{A framework of robust
  transmission design for IRS-aided MISO communications with imperfect cascaded
  channels},'' {\em arXiv preprint arXiv:2001.07054}, 2020.

\bibitem{6587554}
E.~Basar, U.~Aygolu, E.~Panayirci, and H.~V. Poor, ``Orthogonal frequency
  division multiplexing with index modulation,'' {\em IEEE Transactions on
  Signal Processing}, vol.~61, pp.~5536--5549, Nov 2013.

\bibitem{8737925}
S.~{Gopi}, S.~{Kalyani}, and L.~{Hanzo}, ``Coherent and non-coherent multilayer
  index modulation,'' {\em IEEE Access}, vol.~7, pp.~79677--79693, 2019.

\bibitem{meyer2000matrix}
C.~D. Meyer, {\em Matrix analysis and applied linear algebra}, vol.~71.
\newblock {SIAM}, 2000.

\bibitem{deb2012optimization}
K.~Deb, {\em Optimization for engineering design: Algorithms and examples}.
\newblock PHI Learning Pvt. Ltd., 2012.

\bibitem{osborne1969algorithm}
M.~R. Osborne and G.~A. Watson, ``An algorithm for minimax approximation in the
  nonlinear case,'' {\em The Computer Journal}, vol.~12, no.~1, pp.~63--68,
  1969.

\bibitem{7744497}
H.~{Yang}, X.~{Chen}, F.~{Yang}, S.~{Xu}, X.~{Cao}, M.~{Li}, and J.~{Gao},
  ``Design of resistor-loaded reflectarray elements for both amplitude and
  phase control,'' {\em IEEE Antennas and Wireless Propagation Letters},
  vol.~16, pp.~1159--1162, 2017.

\bibitem{van2004optimum}
H.~L. Van~Trees, {\em {Optimum array processing: Part IV of detection,
  estimation, and modulation theory}}.
\newblock John Wiley \& Sons, 2002.

\bibitem{balanis2016antenna}
C.~A. Balanis, {\em Antenna theory: analysis and design}.
\newblock John wiley \& sons, 2016.

\bibitem{342465}
Y.~Pati, R.~Rezaiifar, and P.~Krishnaprasad, ``Orthogonal matching pursuit:
  {Recursive} function approximation with applications to wavelet
  decomposition,'' in {\em {Conference Record of The Twenty-Seventh Asilomar
  Conference on Signals, Systems and Computers, 1993. }}, pp.~40--44 vol.1, Nov
  1993.

\bibitem{needell2009cosamp}
D.~Needell and J.~A. Tropp, ``{CoSaMP}: Iterative signal recovery from
  incomplete and inaccurate samples,'' {\em {Applied and Computational Harmonic
  Analysi}s}, vol.~26, no.~3, pp.~301--321, 2009.

\bibitem{frechet1935generalisation}
M.~Fr{\'e}chet, ``Gen{\'e}ralisation du th{\'e}oreme des probabilit{\'e}s
  totales,'' {\em Fundamenta mathematicae}, vol.~1, no.~25, pp.~379--387, 1935.

\bibitem{wang2019channel}
Z.~Wang, L.~Liu, and S.~Cui, ``Channel estimation for intelligent reflecting
  surface assisted multiuser communications: Framework, algorithms, and
  analysis,'' {\em arXiv preprint arXiv:1912.11783}, 2019.

\bibitem{1614066}
D.~Donoho, ``Compressed sensing,'' {\em IEEE Transactions on information
  theory}, vol.~52, pp.~1289--1306, April 2006.

\bibitem{362841}
{Yuguang Fang}, K.~A. {Loparo}, and {Xiangbo Feng}, ``Inequalities for the
  trace of matrix product,'' {\em IEEE Transactions on Automatic Control},
  vol.~39, pp.~2489--2490, Dec 1994.

\bibitem{380145}
T.~{Eng} and L.~B. {Milstein}, ``Coherent {DS-CDMA} performance in {Nakagami}
  multipath fading,'' {\em IEEE Transactions on Communications}, vol.~43,
  pp.~1134--1143, Feb 1995.

\bibitem{mathai2006generalized}
A.~M. Mathai and R.~K. Saxena, {\em Generalized hypergeometric functions with
  applications in statistics and physical sciences}, vol.~348.
\newblock Springer, 1973.

\bibitem{buchholz2013confluent}
H.~Buchholz, {\em The confluent hypergeometric function: with special emphasis
  on its applications}, vol.~15.
\newblock Springer Science \& Business Media, 1969.

\bibitem{proakis2001digital}
J.~G. Proakis and M.~Salehi, {\em Digital communications}, vol.~4.
\newblock McGraw-hill New York, 2001.

\bibitem{wolfram1}
{Wolfram Research}, ``Complimentary error function.'' [Online]. Available:
  \url{http://functions.wolfram.com/GammaBetaErf/Erfc/26/01/01/}.
\newblock Last visited on 15/2/2020.

\bibitem{gacite}
P.~Sebah and X.~Gourdon, ``Introduction to the {Gamma} function.'' [Online].
  Available: \url{https://www.csie.ntu.edu.tw/~b89089/link/gammaFunction.pdf}.
\newblock Last visited on 15/2/2020.

\bibitem{gradshteyn2014table}
I.~S. Gradshteyn and I.~M. Ryzhik, {\em Table of integrals, series, and
  products}.
\newblock Academic press, 2007.

\bibitem{6389887}
S.~{Kalyani}, ``On {CRB} for parameter estimation in two component gaussian
  mixtures and the impact of misspecification,'' {\em IEEE Transactions on
  Communications}, vol.~60, pp.~3734--3744, December 2012.

\bibitem{wolfram2}
{Wolfram Research}, ``{Gauss hypergeometric function ${}_2F_1$}.'' [Online].
  Available:
  \url{http://functions.wolfram.com/HypergeometricFunctions/Hypergeometric2F1/16/01/01/}.
\newblock Last visited on 15/2/2020.

\bibitem{wolfram3}
{Wolfram Research}, ``{Gauss hypergeometric function ${}_2F_1$}.'' [Online].
  Available:
  \url{http://functions.wolfram.com/HypergeometricFunctions/Hypergeometric2F1/03/06/01/0006/}.
\newblock Last visited on 15/2/2020.

\bibitem{wolfram4}
{Wolfram Research}, ``Incomplete beta function.'' [Online]. Available:
  \url{http://functions.wolfram.com/GammaBetaErf/Beta3/03/01/01/}.
\newblock Last visited on 15/2/2020.

\bibitem{geller1969table}
M.~Geller and E.~W. Ng, ``A table of integrals of the error function.,'' {\em
  Journal of Research of the National Bureau of Standards- {B Mathematical}
  Sciences}, vol.~73B, no.~1, 1969.

\bibitem{xu2009approximation}
R.~Xu, Z.~Zhong, and J.-M. Chen, ``Approximation to the capacity of rician
  fading mimo channels,'' in {\em VTC Spring 2009-IEEE 69th Vehicular
  Technology Conference}, pp.~1--5, IEEE, 2009.

\bibitem{eaton1983multivariate}
M.~L. Eaton, ``Multivariate statistics: a vector space approach.,'' {\em John
  Wiley \& Sons, Inc., 605 Third Avenue., New York, NY 10158, USA, 1983, 512},
  1983.

\bibitem{nakagami1960m}
M.~Nakagami, ``{The m-distribution - A general formula of intensity
  distribution of rapid fading},'' in {\em Statistical methods in radio wave
  propagation}, pp.~3--36, Elsevier, 1960.

\end{thebibliography}
 \bibliographystyle{ieeetr} 

\end{document}